%% file: KMNOO20230629arXiv.tex
\documentclass[12pt,reqno]{amsart}
\usepackage{amsmath,amssymb}
\usepackage{ascmac,fancybox}
\usepackage{amscd}
\usepackage[dvipdfmx]{graphicx}
\usepackage[all]{xy}
\usepackage{here}
\usepackage{color}
\usepackage {booktabs}

\makeatletter
\@addtoreset{equation}{section}
\makeatother
\usepackage{enumerate}

\input{define.tex}

\begin{document}


\title{A novel symmetry in nanocarbons:
pre-constant discrete principal curvature structure}

\author{Yutaro Kabata, Shigeki Matsutani, Yusuke Noda, Yuta Ogata, Jun Onoe }

\date{\today}

\begin{abstract}
Since the first-principles calculations in quantum chemistry precisely provide possible configurations of carbon atoms in nanocarbons, we have analyzed the geometrical structure of the possible carbon configurations and found that there exists a novel symmetry in the nanocarbons, i.e., the pre-constant discrete principal curvature (pCDPC) structure. 
In terms of the discrete principal curvature based on the discrete geometry for trivalent oriented graphs developed by Kotani, Naito, and Omori [Comput. Aided Geom. Design, {\bf{58}}, (2017), 24-54], we numerically investigated discrete principal curvature distribution of the nanocarbons, C$_{60}$, carbon nanotubes, C$_{120}$ (C$_{60}$ dimer), and C$_{60}$-polymers (peanut-shaped fullerene polymers).
While the C$_{60}$ and nanotubes have the constant discrete principal curvature (CDPC) as we expected, it is interesting to note that the C$_{60}$-polymers and C$_{60}$ dimer also have the almost constant discrete principal curvature,  i.e., pCDPC, which is surprising.
A nontrivial pCDPC structure with revolutionary symmetry is available due to discreteness, though it has been overlooked in geometry.
In discrete geometry, there appears a center axisoid which is the discrete analogue of the center axis in the continuum differential geometry but has three-dimensional structure rather than a one-dimensional curve due to its discrete nature. 
We demonstrated that such pCDPC structure exists in nature, namely in the C$_{60}$-polymers.
Furthermore, since we found that there is a positive correlation between the degree of the CDPC structure and stability of the configurations for certain class of the C$_{60}$-polymers, we also revealed the origin of the pCDPC structure from an aspect of materials science.

\end{abstract}

\maketitle

\section{Introduction}
Interdisciplinary field between mathematics and materials science has recently been attracted.
Especially geometrical properties of carbon networks determine their material properties \cite{ShimaOnoe}.
The relations between carbon networks and discrete differential geometry are currently concerned after Terrones and Mackay proposed a carbon network, which is supposed to be a discrete version of the Schwarzian surface, a triply periodic minimal surface with negative Gaussian curvature; it is known as the classical Mackay crystal \cite{MT, T}.
Further $K_4$ crystal was theoretically proposed based on mathematical consideration, and its properties were studied by the first-principles calculations \cite{IKNSKA}, though it is unstable from the viewpoint of phonon dispersion \cite{YTSKMI}.
Recently Kotani, Naito and Omori developed excellent tools in the discrete differential geometry for trivalent graphs embedded in three-dimensional euclidean space $\RR^3$, and investigated the carbon networks in terms of the discrete differential geometry \cite{KNO, KNT}.
Furthermore, Sunada also described mathematical structures behind them \cite{S}.
However, there have been no geometrical analyses of the actual nanocarbon geometry obtained by the first-principles calculations as far as we know.

In this paper, we have investigated the discrete differential geometrical properties of nanocarbons, mainly C$_{60}$-polymers extended to the one-dimensional direction due to the linkage among C$_{60}$s, by using the data of the actual nanocarbon geometry obtained by first-principles calculations.
Such an analysis is the first attempt as far as we know.

The C$_{60}$-polymers are sometimes referred to as the peanut-shaped fullerene polymer (PSFP) due to their shape.
The C$_{60}$-polymer were discovered in 2003 \cite{ONAH,UONIO}, which have been produced by electron beam irradiation of C$_{60}$  thin films \cite{ONAH}.
The electron system on the C$_{60}$-polymer exhibits Tomonaga-Luttinger liquid states, which has a novel behavior due to the submanifold quantum system different from conventional nanocarbons such as nanocarbons, nanotubes, and graphene \cite{OISYK}, and thus it is fascinating to understand its geometrical property but it has never been revealed yet. 
Since the atomic configurations of the C$_{60}$-polymers were determined by the first principles calculations \cite{BO, NOO}, we investigate the differential geometrical properties of the C$_{60}$-polymers by analyzing the atomic configuration data in terms of the discrete geometrical tools introduced by Kotani, Naito and Omori \cite{KNO}.
Such a discrete differential geometrical analysis on the atomic configurations given by the first principles calculations is crucial and novel so far.

As shown in Figure \ref{fg:PSFP1}, the carbon atoms in the C60-polymer look like the boundary surface of a solid of revolution. 
It may be decomposed to three regions from a continuum geometrical viewpoint, i.e., the elliptic region (positive Gauss curvature), the parabolic region (null-Gauss curvature), and the hyperbolic (negative Gauss curvature).
The shape might be similar to the Delaunay surface, or a constant mean curvature (CMC) surface.
Though there have been several reports on the relation between carbon networks (or nanocarbons) and the Delaunay surfaces, \cite{Vassilev,Xie_etal}, our numerical analysis showed that the shape of the C$_{60}$-polymer cannot be regarded as the Delaunay surface as will be mentioned in this paper.
However, instead of the CMC property, the geometrical analysis of the first-principles computational results leads to a novel property that shows almost constant discrete principal curvature due to its discrete nature; we refer to it as \emph{pre-constant discrete principal curvature (pCDPC)}.

To the best of our knowledge, the pCDPC surfaces have never been mentioned in any literature so far. 
It is a novel symmetry of carbon atom networks, which has neither been investigated anyplace.

Hence, we have further investigated the geometrical properties of the typical kinds of the nanocarbons, i.e., C$_{60}$, C$_{60}$-dimer, nanotubes the C$_{60}$-polymers in terms of tools in \cite{KNO}, and found that the novel geometrical structure, the pCDPC, also holds in all of those nanocarbons. 
This finding is interesting since those nanocarbons have the pCDPC, which has never been described. 
In other words, the geometrical properties can be regarded as a novel discrete geometrical symmetry hidden in the nanocarbons. It is emphasized that this symmetry was first discovered by an interdisciplinary study between geometry and materials science in this paper. 
We report this symmetry in this paper.

The contents of this paper is following.
Section \ref{sec:S1} is a mathematical preliminary to show the discrete geometrical properties following Kotani, Naito, and Omori \cite{KNO, KMO}.
We review the shapes of the C$_{60}$-polymers obtained using the first-principles calculations as in \cite{NOO} in Section \ref{sec:S2}.
By the discrete geometrical analysis, the geometrical properties of the C$_{60}$-polymers, and C$_{60}$-dimer (C$_{120}$), and nanotubes are shown in Sections \ref{sec:S3} and \ref{sec:S4}.
Thereafter, we discuss these results in Section \ref{sec:S5}.
Finally, we describe the conclusions and perspectives in the present study in Section \ref{sec:S7}.

\section{Mathematical preliminary}\label{sec:S1}

In this section, we review the discrete differential geometry of trivalent oriented graphs embedded into the three-dimensional euclidean space $\RR^3$ following Kotani, Naito and Omori \cite{KNO}.
We refer to it as KNO construction, and based on it, we introduce the discrete principal curvature as follows \cite{KMO}:

Let $X=(V_X, E_X)$ be a trivalent oriented graph;
for each vertex $x \in V_X$, we have the subset of edges $E_x=\{e_1, e_2, e_3\}\subset E_X$ whose incident point is $x$, i.e., $x=t(e_i)$.
The orientation is determined so that each cycle in $X$ is consistent.

We consider the embedding $\iota: X \to \RR^3$; we denote $\iota(x)$ by $\bx$ and $\iota(e)$ by $\be$ for $x\in V_X$ and $e \in E_X$.
The tangent plane $T_{\bx}$ with the normal vector $\bn_x \in \RR^3$ at $\bx$ is defined by 
$$
\bn_x =\frac{(\be_1-\be_3)\wedge(\be_2-\be_3)}
{|(\be_1-\be_3)\wedge(\be_2-\be_3)|}
= \frac{\be_1\wedge\be_2+\be_2\wedge\be_3+\be_3\wedge\be_1}
{|\be_1\wedge\be_2+\be_2\wedge\be_3+\be_3\wedge\be_1|}
$$
for $\be_i=\iota(e_i)$ of $E_x=\{e_1, e_2, e_3\}$.
For $x \in V_X$ and $E_x=\{e_1, e_2, e_3\}\subset E_X$, let each adjacent vertex be denoted by $x_i$, i.e., $e_i=(x, x_i)$, $(i =1, 2, 3)$ so that the ordered vertices $x_1,x_2,x_3$ form an oriented triangle $\Delta_x$ given by $\bx_i=\iota(x_i)$, ($i=1, 2, 3$), i.e., $\Delta_x =(\bx_1,\bx_2,\bx_3)\subset \RR^3$.
For each $\bx_i$, the normal vector $\bn_i=\bn_{x_i}$, $(i=1,2,3)$ is assigned.
We set $\bv_1=\be_1-\be_3=\bx_1-\bx_3$ and $\bv_2=\be_2-\be_3=\bx_2-\bx_3$. 
Let the euclidean inner product be $\langle, \rangle: \RR^3 \times \RR^3 \to \RR$.
We introduce the discrete first fundamental form $I_x$ defined by
$$
I_x =(g_{ij}) =
\begin{pmatrix}
\langle \bv_1, \bv_1 \rangle &
\langle \bv_1, \bv_2 \rangle \\
\langle \bv_2, \bv_1 \rangle &
\langle \bv_2, \bv_2 \rangle \\
\end{pmatrix}.
$$
We also define the directional derivative $\nabla_i \bn_x$ of the normal vector $\bn$ for the vector $\bv_i$ by
$$
\nabla_i \bn_x = \mathrm{Proj}_x[\bn_i -\bn_3] =(\bn_i -\bn_3) -
\langle \bn_i -\bn_3,\bn_x\rangle \bn_x
$$
for $i =1, 2$; we note that $\mathrm{Proj}_x$ exhibits the orthogonal projection onto the tangent plane $T_{\bx}$.
The discrete second fundamental form $II_x$ is defined by
\begin{gather*}
\begin{split}
II_x &=
-\begin{pmatrix}
\langle \bv_1,\nabla_1 \bn_x\rangle &
\langle \bv_1,\nabla_2 \bn_x\rangle \\
\langle \bv_2,\nabla_1 \bn_x\rangle &
\langle \bv_2,\nabla_2 \bn_x\rangle \\
\end{pmatrix}\\
&=
-
\begin{pmatrix}
\langle \bv_1,  \bn_1- \bn_3\rangle &
\langle \bv_1,  \bn_2- \bn_3\rangle \\
\langle \bv_2,  \bn_1- \bn_3\rangle &
\langle \bv_2,  \bn_2- \bn_3\rangle \\
\end{pmatrix}.
\end{split}
\end{gather*}
The discrete Weingarten map is introduced by $\Gamma_x=I_x^{-1} II_x$.
Using $\Gamma_x$, 
the discrete mean curvature $H_x$ and the discrete Gaussian curvature $G_x$ are defined by
$$
H_x = \mathrm{tr}(\Gamma_x), \quad
G_x = \mathrm{det}(\Gamma_x).
$$
Based on the KNO construction \cite{KNO}, the discrete principal curvatures $k_1$ and $k_2$ are defined by,
$$
(t-k_1)(t-k_2)=t^2-H_x t + G_x,
$$
such that $|k_1|\ge |k_2|$ \cite{KMO}.
They are natural discrete analogue of the principal curvature in the differential geometry, whose inverse should be regarded as the discrete analogue of the curvature radius, i.e., $\rho_a= 1/k_a$ $(a=1,2)$.

We define the center of the curvature radius $\bc_{x,a}$ by
\begin{equation}
     \bc_{x,a}= \bx - \rho_a \bn_x.
\label{eq:CAxisoid}
\end{equation}
Even though the solid of revolution has the center axis, for the discrete analogue $\iota(X)$ of the solid of revolution, the set of $\{\bc_{x,1}\ |\ x\in V_X\}$ does not exist on a straight line, in general.
We, thus, refer to the set  as the \emph{center-axisoid} \cite{KMO}.
The distribution of the center-axisoid in three dimensional space shows the nature of the discrete geometry.

\section{Shapes of the C$_{60}$-polymer obtained using the first-principles calculation}\label{sec:S2}

\begin{figure}[H]
\begin{center}
\includegraphics[width=0.55\hsize,bb=0 0 1117 489]{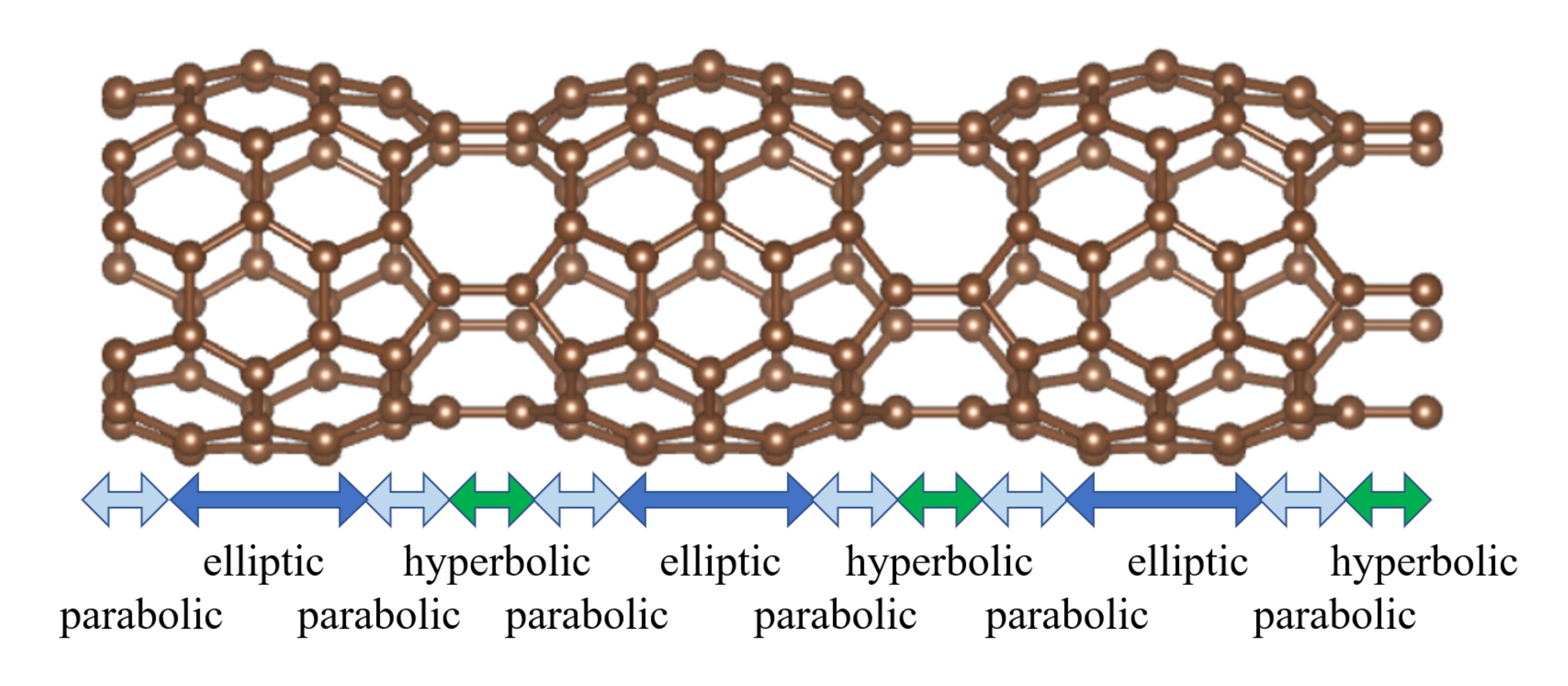}

\end{center}

\caption{
The C$_{60}$-polymers by the first-principles  calculation:
FP5N \cite{NOO}
}\label{fg:PSFP1}
\end{figure}

In the present work, we have mainly investigated the C$_{60}$-polymer, which is extended to the one-dimensional direction due to the linkage among C$_{60}$s, illustrated in Figure \ref{fg:PSFP1}.
The C$_{60}$-polymers are produced by irradiating C$_{60}$ thin films with electron beams \cite{ONAH}. 
The shape of the C$_{60}$-polymers has been confirmed by analysis of infrared spectra in combination with first-principles calculations \cite{TON}, and of the electron diffraction patterns by the high-resolution transmission electron microscopy and the high-resolution cryo-transmission electron microscopy \cite{MOY, MYO}, but the atomic configurations of its geometrical shape are not still determined precisely.
Thus we investigate the shapes determined by the first-principles  calculation.
We analyzed them from a geometrical viewpoint using the tools prepared in the previous section.

However there are several local minimums of the shapes argued in \cite{WH,NO,NOO}.
We have mainly analyzed the shape of FP5N, which has the lowest total energy in the computation but also gave some results for two shapes as references; (later we will consider further 13 shapes in Section \ref{sec:S5}.)

We mention the first-principles calculations in the present work following \cite{NOO}.

The authors in \cite{NOO} used a generalized Stone-Wales (GSW) transformation to search the smaller energy of the configurations of carbon atoms for the C$_{60}$-polymers, to which are referred as the PSFP models.
They compared their results with those of T3 model of the C$_{60}$-polymer  proposed by Wang and Huang \cite{WH}. 
We also regard it as one of the PSFP models. 

Following them, we performed the first-principles calculations in the present work.
In the present calculations, we employed Vienna ab-initio simulation package (VASP) \cite{KF} based on density functional theory (DFT) \cite{HK} to optimize  the  coordinates of carbon atoms and the volume of a periodic unit cell, and to investigate the energetic properties of the PSFP models. 
In our DFT calculations, a generalized gradient approximation (GGA) exchange-correlation functional proposed by Perdew, Burke, and Ernzerhof (PBE) \cite{PBE} and a projector augmented wave (PAW) pseudopotential were used \cite{KJ}. 
The three-dimensional periodic unit cell of the rectangular cuboid shape with $L \times 20.000 \times 20.000$ \AA${}^3$ was used, where $L$ is a lattice parameter of the primitive C$_{60}$-polymer region and the $L$ value for each PSFP model is listed in Table \ref{2tb:Ex50}. 
A cutoff energy for the plane-wave basis set was set to $500$ eV, and a $10\times 1\times 1$ $\Gamma$-centered $k$-point grid was used for the geometrical optimization of the PSFP models. The atomic coordinates and the lattice parameters $L$ were optimized with the total energy convergence criterion being $10^{-8}$ eV between two ionic optimization steps.

\begin{figure}[H]
\begin{center}
\includegraphics[width=0.30\hsize,bb= 0 0 1224 790]{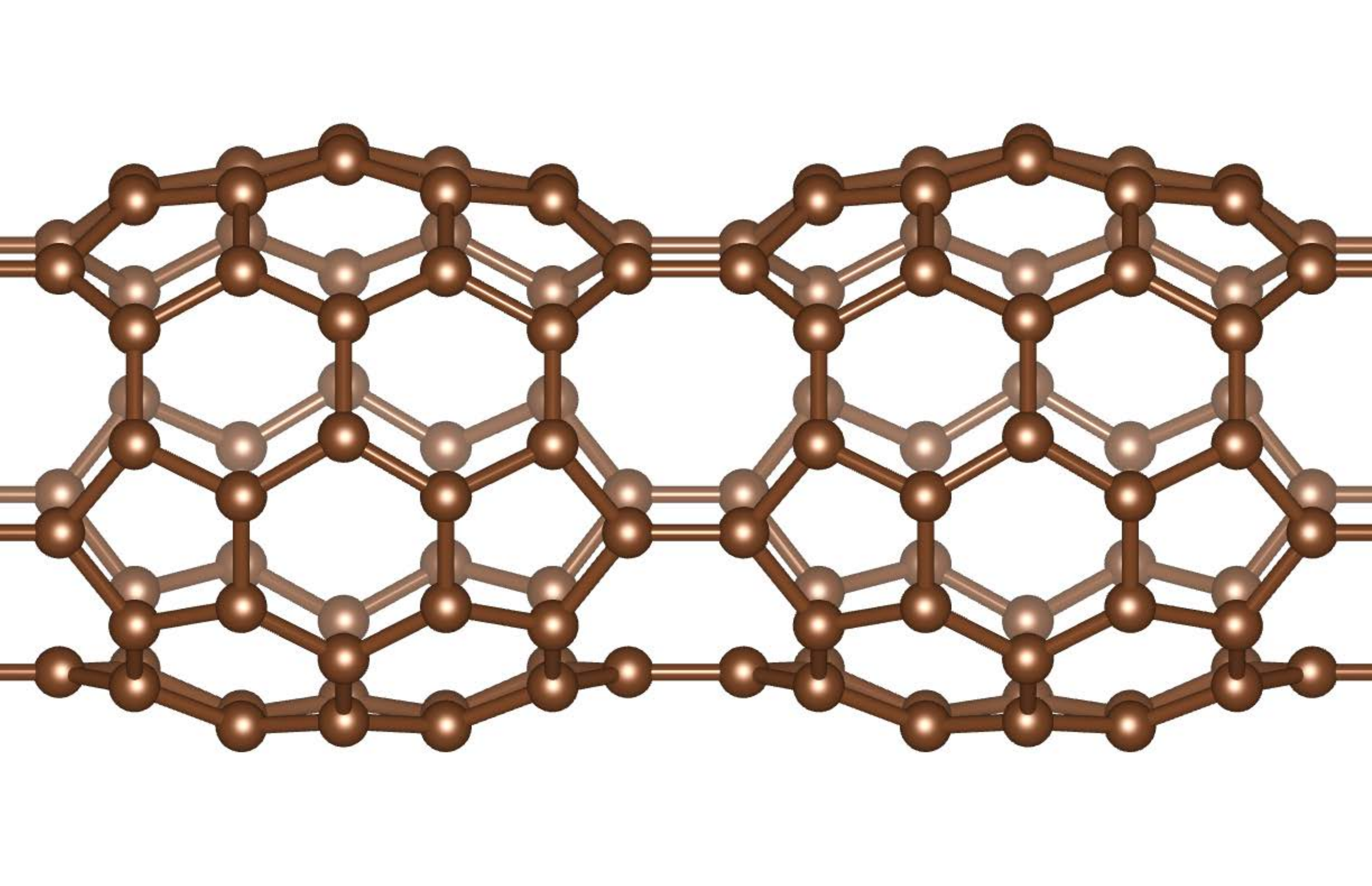}
\includegraphics[width=0.30\hsize,bb= 0 0 1224 790]{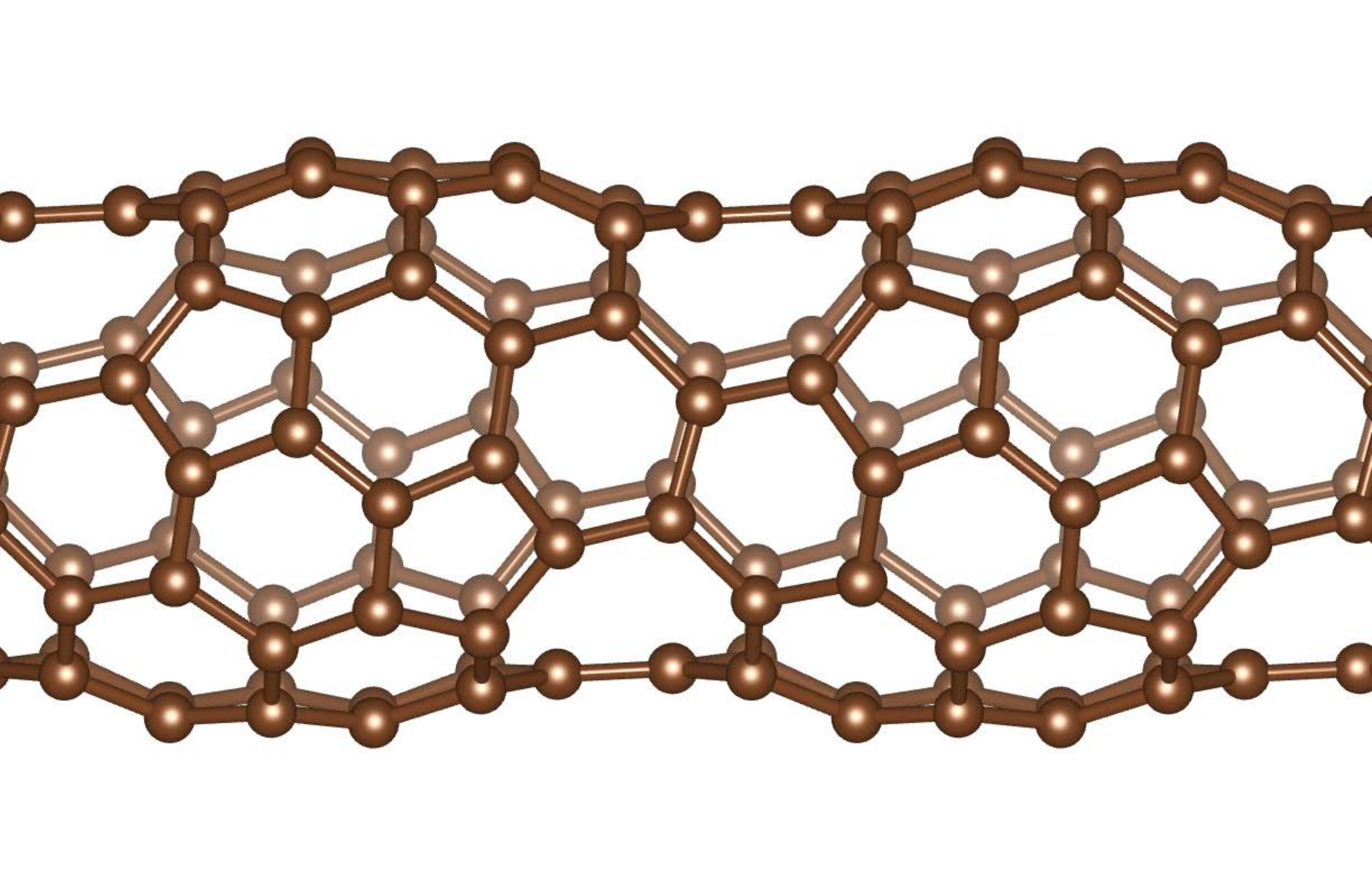}

 (a)
\hskip 0.30\hsize (b)

\includegraphics[width=0.30\hsize,bb= 0 0 1224 790]{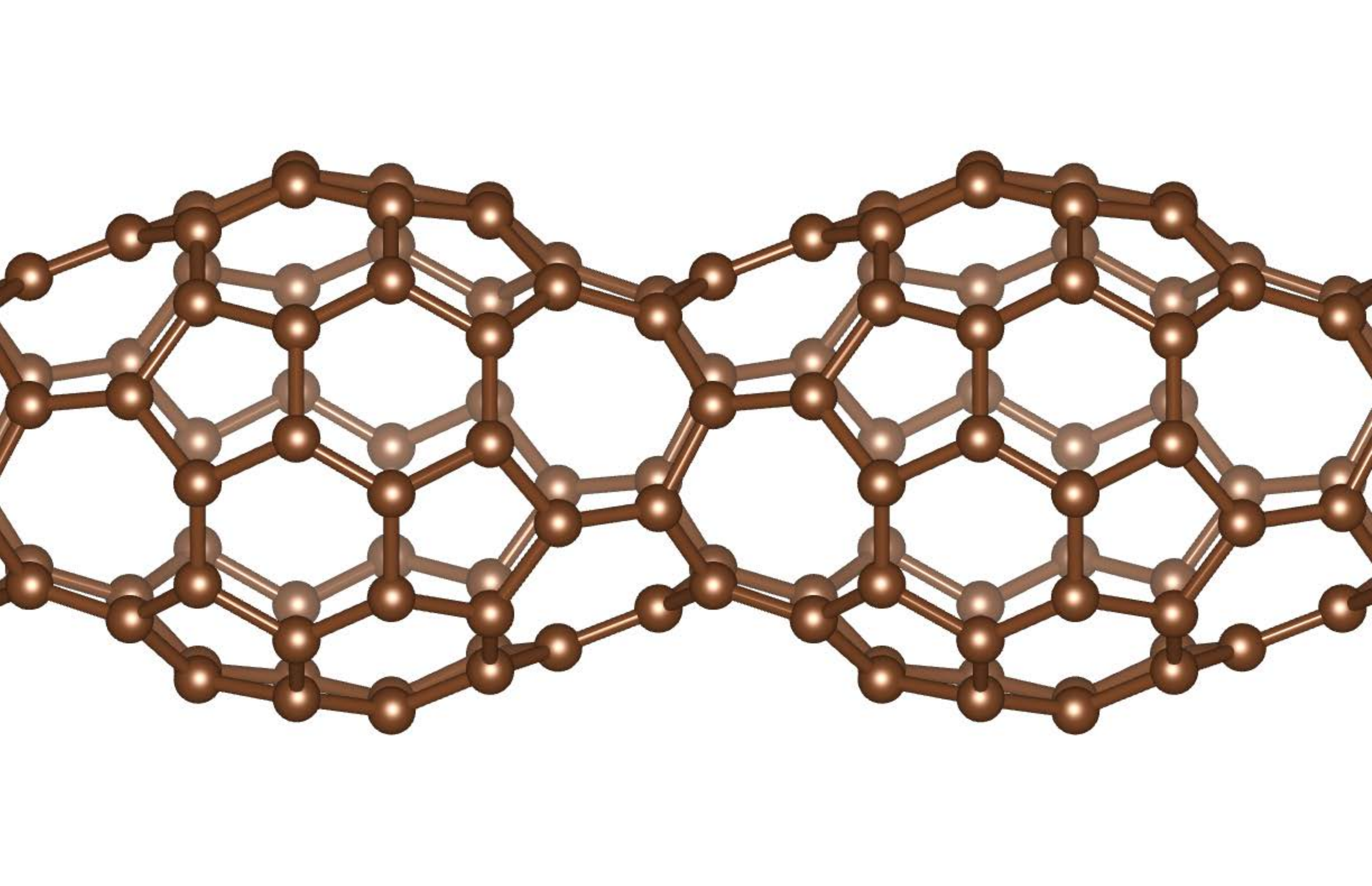}

(c)

\caption{
The shape of the C$_{60}$ polymers given by the first-principles calculations I:FP5N (a),  FP4L (b), and T3 (p).
}\label{fg:C60polymer1}

\end{center}

\end{figure}

The calculated results for three of them are summarized in Table \ref{2tb:Ex50}, and their shapes are displayed in Figure \ref{fg:C60polymer1}.

\begin{table}[htb]
\centering
\caption{Computational results of the C$_{60}$-polymers \cite{NOO}:
$G$, $L$, $E_{\mathrm{tot,uc}}$, $E_{\mathrm{tot,a}}$, and $E_g$ 
 stand for a symbol of point group symmetry, lattice parameter of the C$_{60}$-polymer region, total energy per unit cell, total energy per atom, and electronic band gap between valence band maximum and conduction band minimum. The total energy values indicate the energy difference from the total energy of the T3 model.}\label{2tb:Ex50}
  \begin{tabular}{c|ccccc}
PSFP&  $G$& $L$& $E_{\mathrm{tot,uc}}$ 
& $E_{\mathrm{tot,a}}$ &$E_g$\\
& & (\AA)&  (eV/cell) & (eV/atom) & (eV)\\
\midrule
FP5N & $D_{5h}$ & 8.128  & -2.177  & -0.0363 & \\
FP4L & $C_{2h}$ & 8.250  & -1.233  & -0.0206 & 0.026\\
T3 &   $D_{5d}$ & 8.694  & 0.000  & 0.0000&1.169 \\
\bottomrule
  \end{tabular}

\end{table}

\section{Geometrical properties of the C$_{60}$-polymers}\label{sec:S3}

\begin{figure}[H]
\begin{center}
\includegraphics[width=0.35\hsize, bb=0 0 579 526]{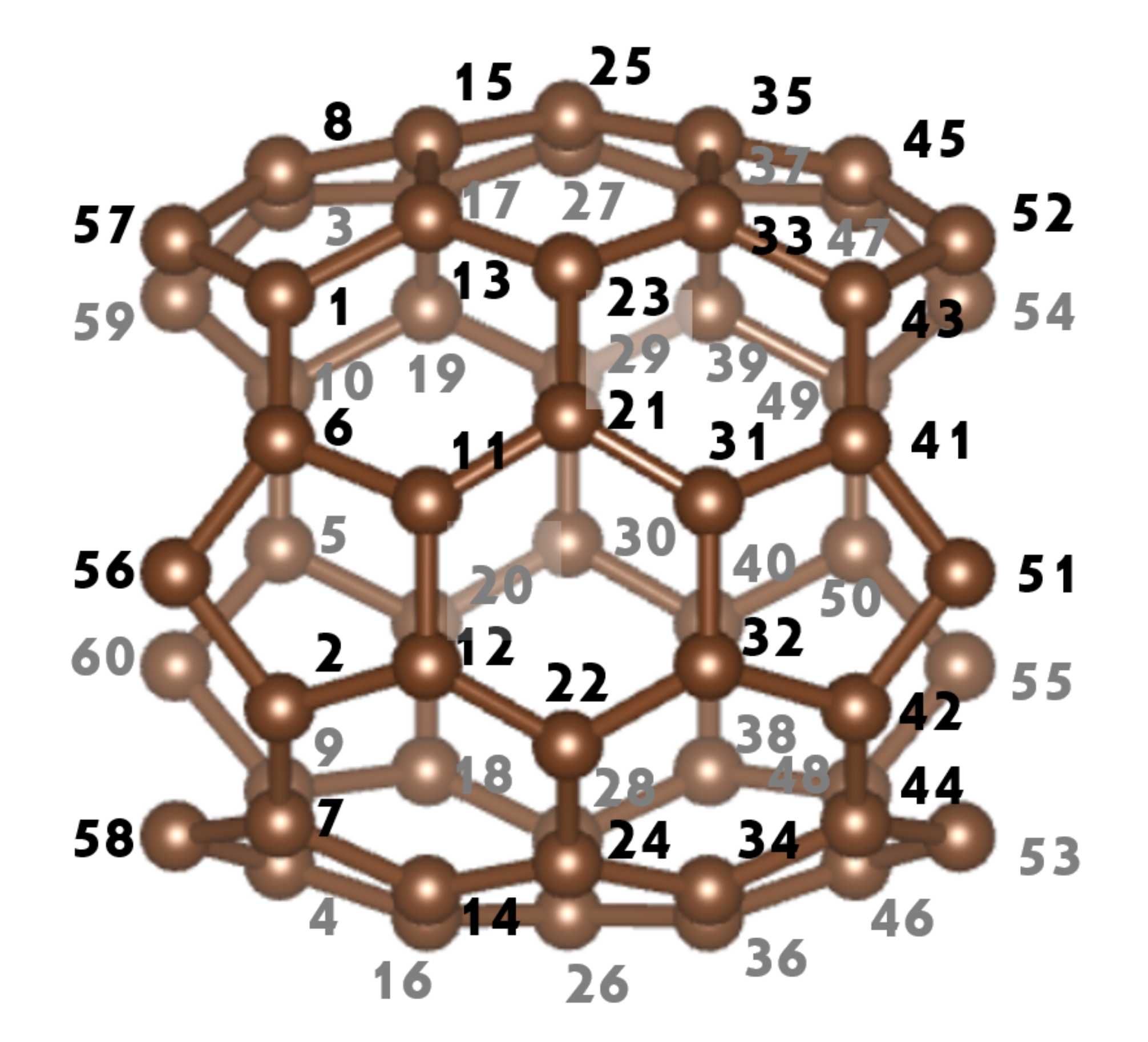}

 (a)

\includegraphics[width=0.50\hsize, bb=0 0 523 300]{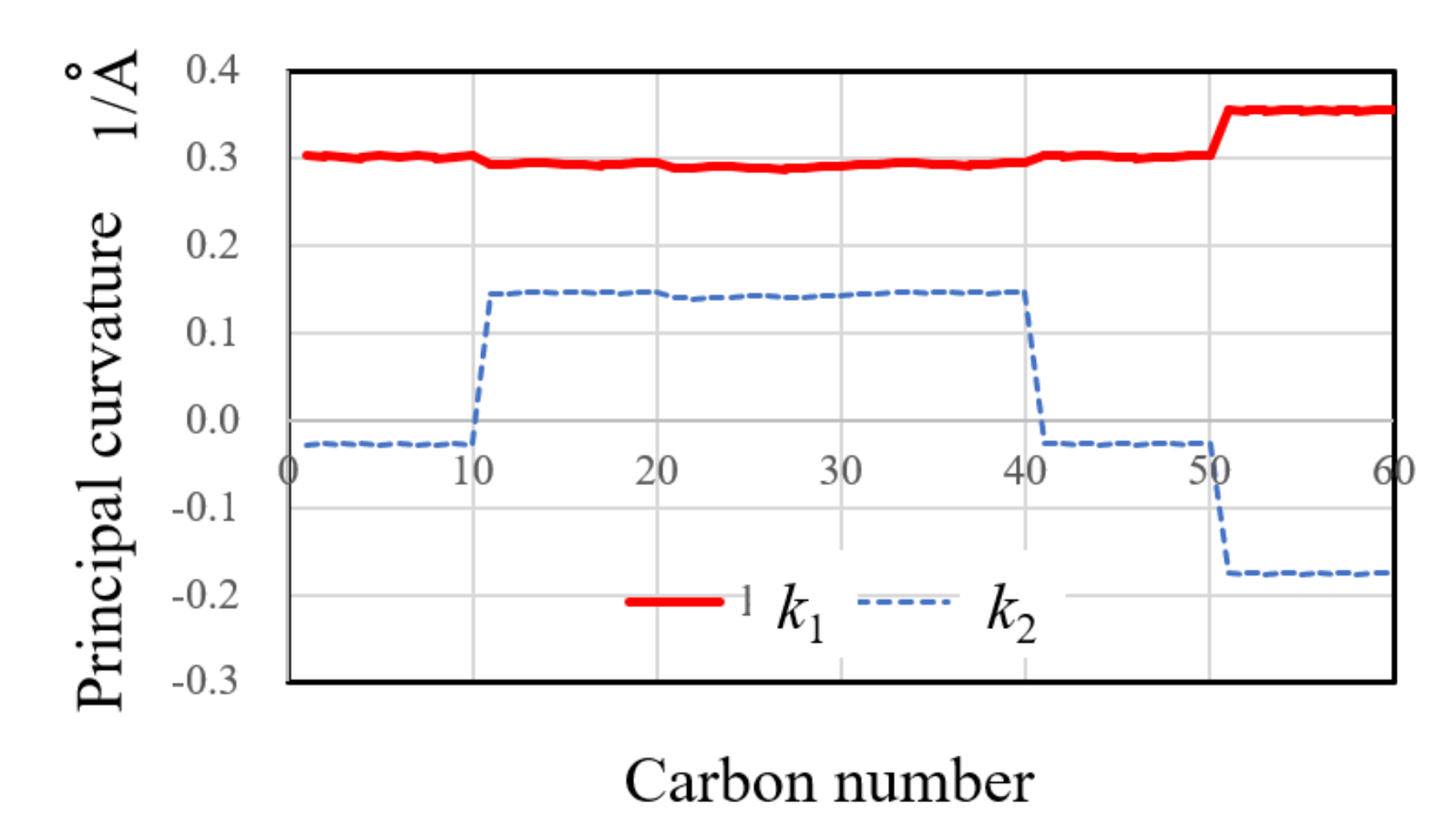}

 (b)

\includegraphics[width=0.50\hsize, bb=0 0 530 293]{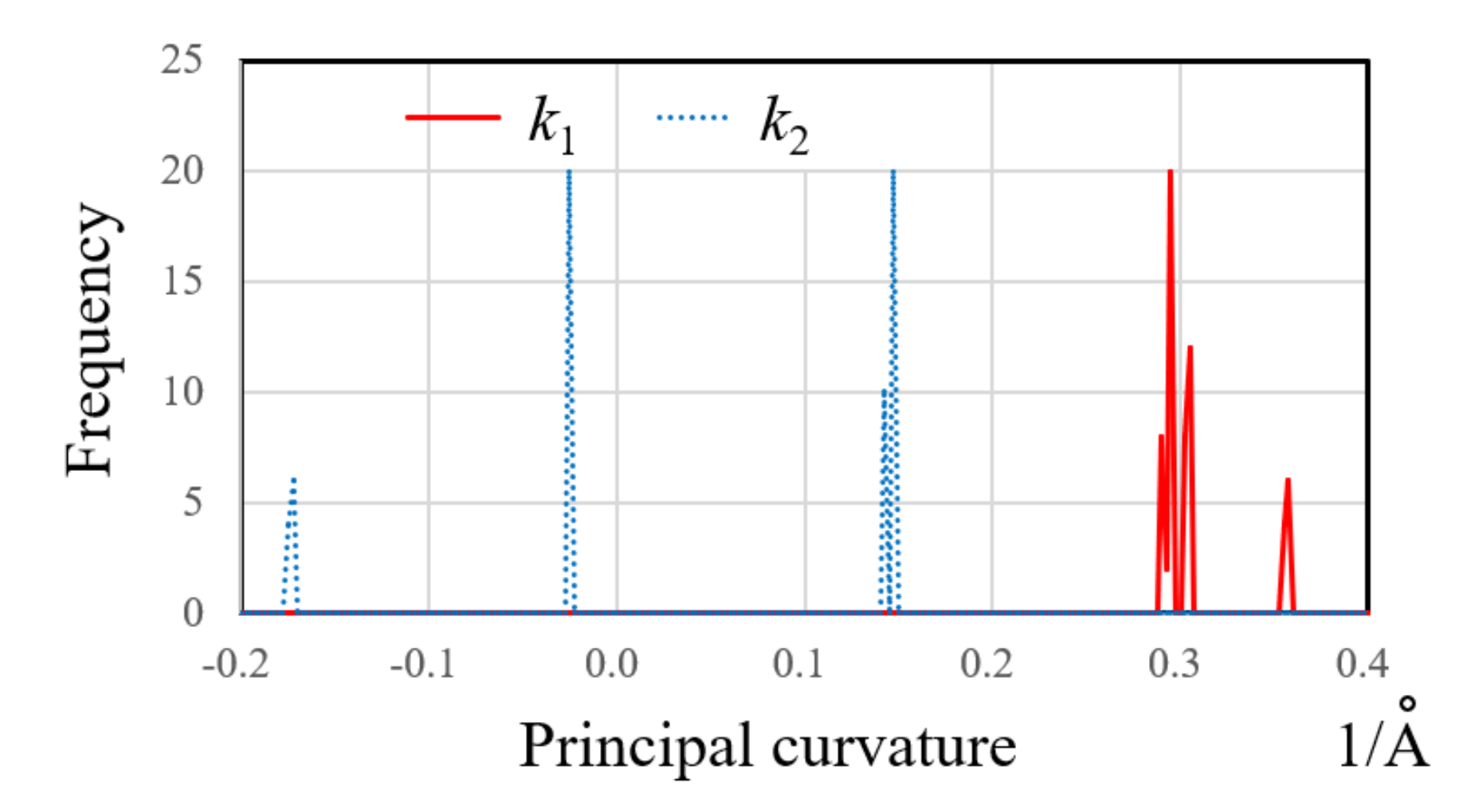}

(c)

\caption{
The discrete principal curvature distribution of the C$_{60}$-polymer:
(a) shows the numbering of the carbon atoms in the C$_{60}$-polymer (FP5N), (b) the value of the discrete principal curvatures $k_a$ $(a=1,2)$ at each carbon atom, and (c) the frequency.
}\label{fg:FP5N_dist}

\end{center}

\end{figure}

As illustrated in Figure \ref{fg:PSFP1}, we may observe that the shape of the C$_{60}$-polymers has the three regions from a continuum geometrical viewpoint; the elliptic region (positive Gauss curvature), the parabolic region (null-Gauss curvature), and the hyperbolic region (negative Gauss curvature).
The discrete Gauss curvature $G_x$ is also given by the product of the principal curvatures $k_1$ and $k_2$, i.e., $G_x= k_1 k_2$.
We geometrically analyzed the configuration data obtained by the first-principles calculations by using the geometrical tools described in Section \ref{sec:S1}.
We obtained the distributions of the discrete principal curvature of the configuration of the C$_{60}$-polymers as trivalent regular graphs.
Figure \ref{fg:FP5N_dist}  displays (a) the numbering of the carbons in the C$_{60}$-polymer, (b) the distribution of the discrete principal curvature at each carbon, and  the frequency of the discrete principal curvatures $k_1$ and $k_2$.
As we predicted above, the distribution of the discrete principal curvature shows  that the region of the distribution is decomposed into the three regions, $k_1 k_2 >0$ (elliptic), $k_1 k_2 \sim 0$ (parabolic), and $k_1 k_2 <0$ (hyperbolic ).
However it is remarked that the first principal curvature $k_1$ is almost constant, which is supported by the frequency of the discrete principal curvatures $k_1$ and $k_2$ in Figure \ref{fg:FP5N_dist} (c).

\begin{figure}[H]
\begin{center}
\includegraphics[width=0.35\hsize, bb=0 0 564 262]{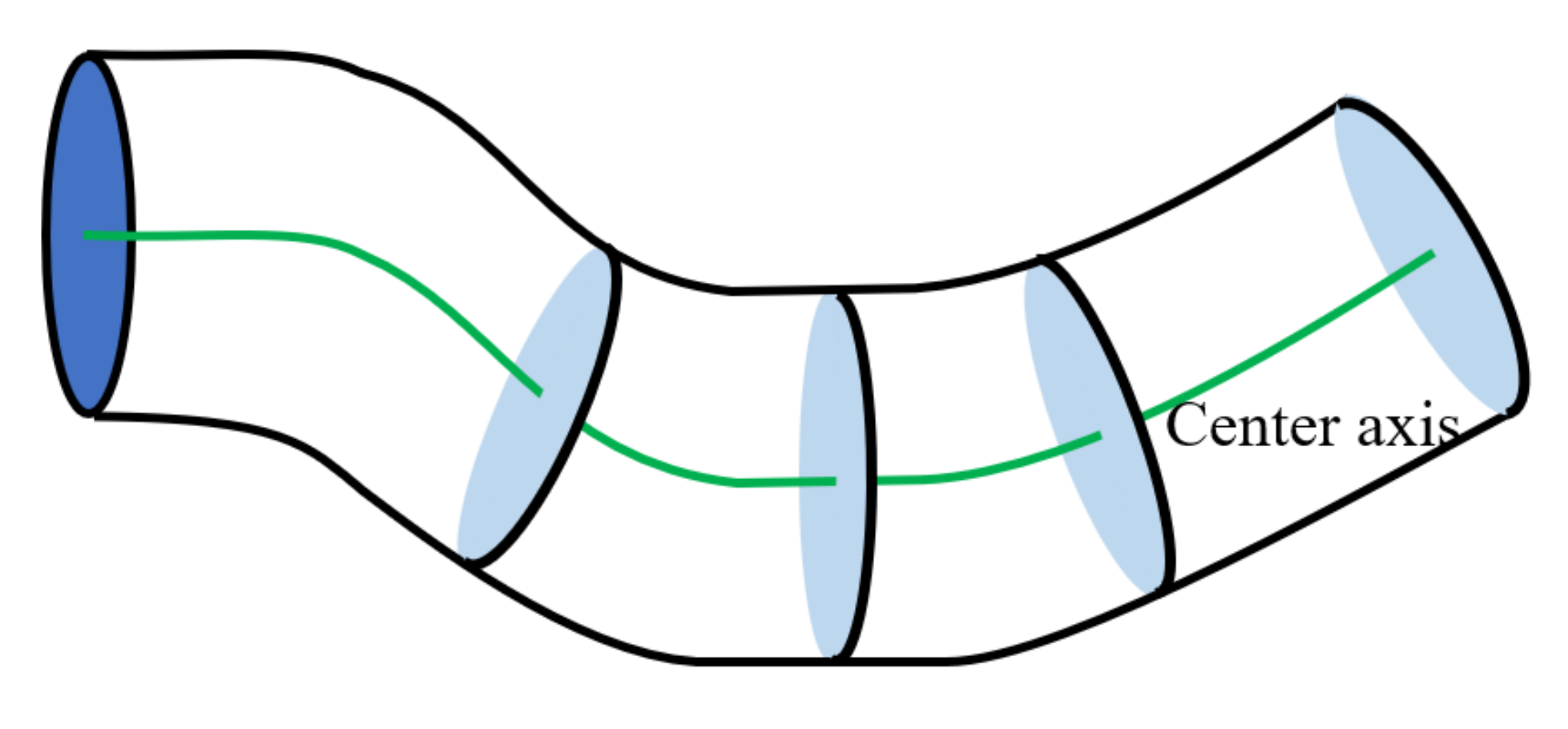}
\includegraphics[width=0.35\hsize, bb=0 0 513 210]{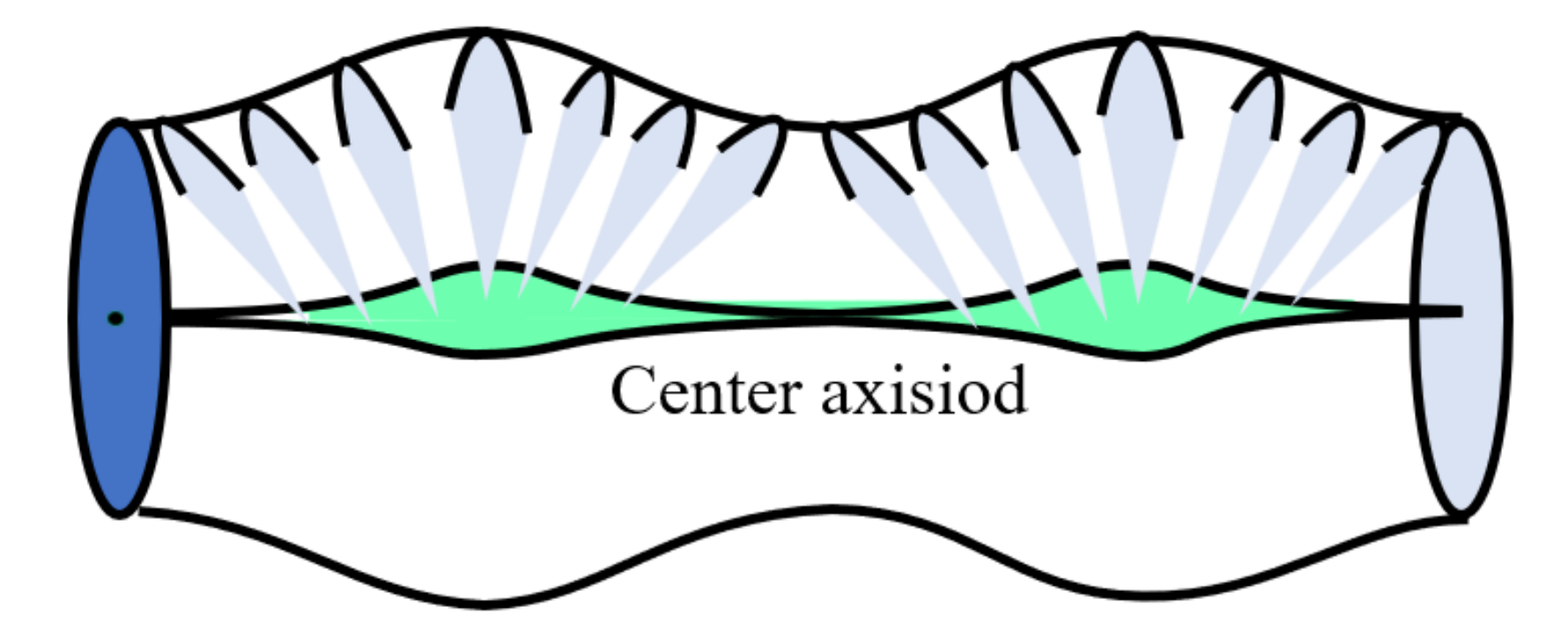}

\hskip 0.03\hsize (a)
\hskip 0.35\hsize (b)

\caption{
Non-trivial CPC and CDPC surfaces:
(a) is a CPC surface and (b) is a CDPC surface.
}\label{fg:CPC_CDPC}

\end{center}

\end{figure}

In differential geometry (as continuum theory), 
it is so strong that the restriction of the principal curvature of surfaces to be constant \cite{KMO}. 
The examples of the constant principal curvature surfaces (CPC surfaces)  in differential geometry are spheres and cylinders, and non-trivial cases are illustrated in Figure \ref{fg:CPC_CDPC} (a).
This means that the surface of revolution with CPC structures must be cylinders or spheres only in the continuum geometry.

\begin{figure}[H]
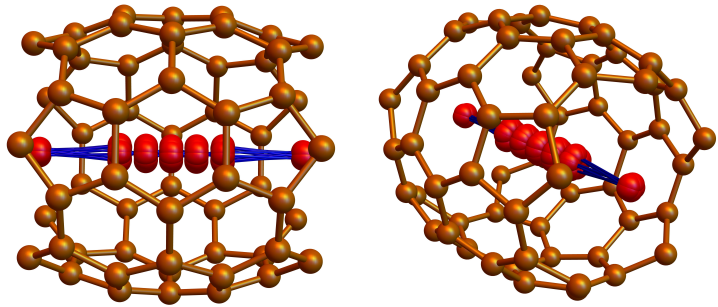

\begin{center}
\includegraphics[width=0.30\hsize, bb=0 0 5000 5125]{centeraxisiod1.pdf}
\includegraphics[width=0.30\hsize, bb=0 0 5000 4915]{centeraxisiod2.pdf}

\caption{
Center axisoid of the C$_{60}$-polymer (FP5N):
The carbon atom configuration of FP5N and its center axisoid $\{\bc_{x,1}\}$ given by (\ref{eq:CAxisoid}) are illustrated.
Red points show $\{\bc_{x,1}\}$.
}\label{fg:CAxisoid}

\end{center}

\end{figure}

However, we encounter a different aspect in the discrete geometry.
Due to the discrete nature, we may find further non-trivial constant discrete principal curvature (CDPC) surfaces displayed in Figure \ref{fg:CPC_CDPC} (b), since the discrete principal curvature is determined locally and the center-axisoid has a nontrivial structure \cite{KMO}.
There appears inconsistency of the center axis of the surfaces of revolution in the discrete versions, which was already remarked by Kotani, Naito and Omori \cite{KNO} for a nanotube.
The structure of the C$_{60}$-polymer is surely determined by such a discrete nature.
We compute the center axisoid of the C$_{60}$-polymer using the formula (\ref{eq:CAxisoid}).
Figure \ref{fg:CAxisoid} displays its distributions of the C$_{60}$-polymer; we found its three dimensional structure rather than one-dimensional line.
Even though its shape is not CDPC precisely, the first principal curvature in the C$_{60}$-polymer of FP5N is almost constant.
The existence of the center axisoid provides such co-existence of non-trivial shape of discrete analogue the surfaces of revolution and the almost CPC structure.
Thus we refer to the discrete surface as the \emph{pre-constant discrete principal curvature surface (pCDPC surface)} as mentioned in Introduction.

\section{Geometrical properties of the nanocarbons}\label{sec:S4}

\begin{figure}[H]
\begin{center}
\includegraphics[width=0.26\hsize, bb=0 0 321 319]{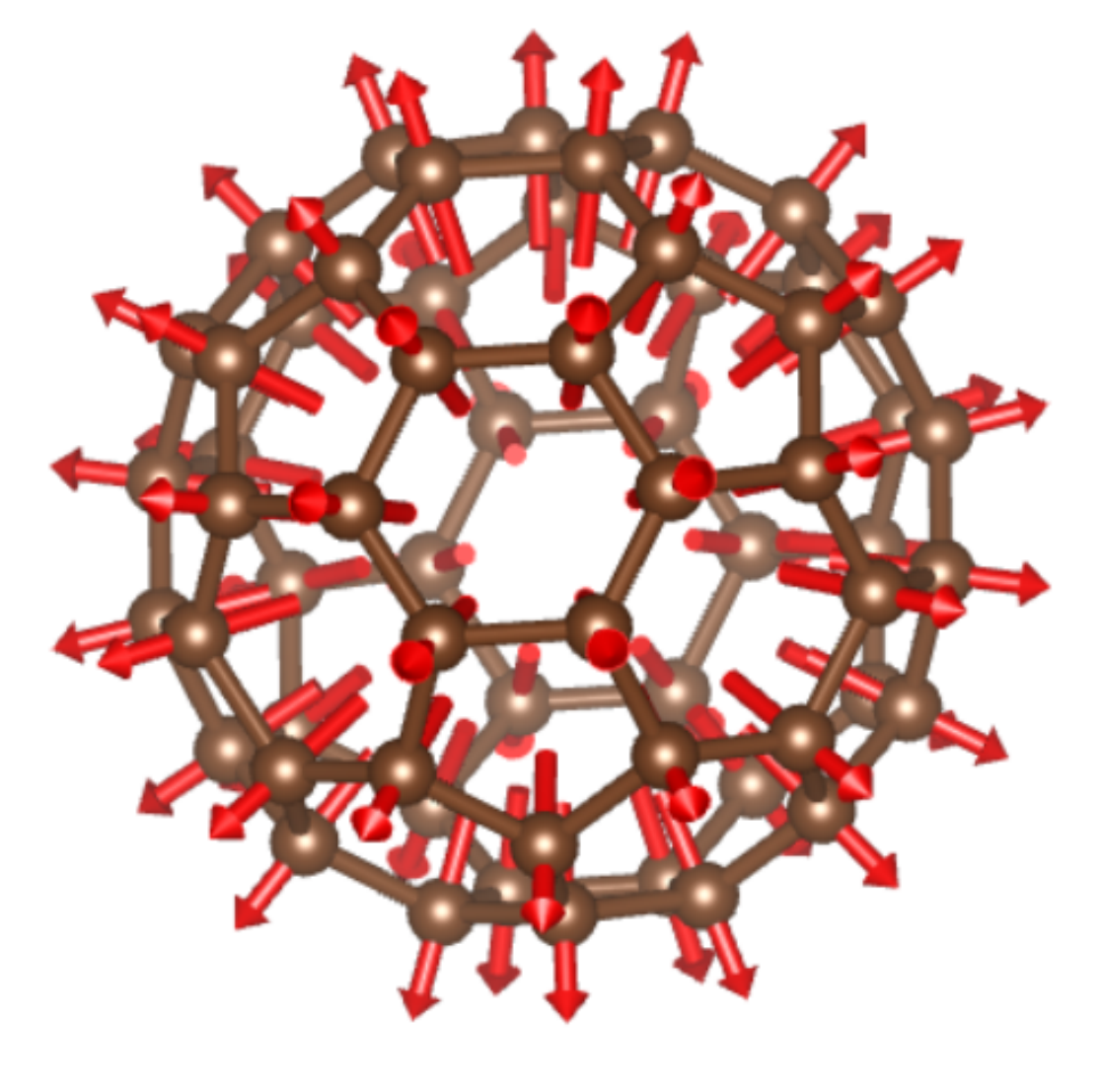}
\includegraphics[width=0.22\hsize, bb=0 0 297 338]{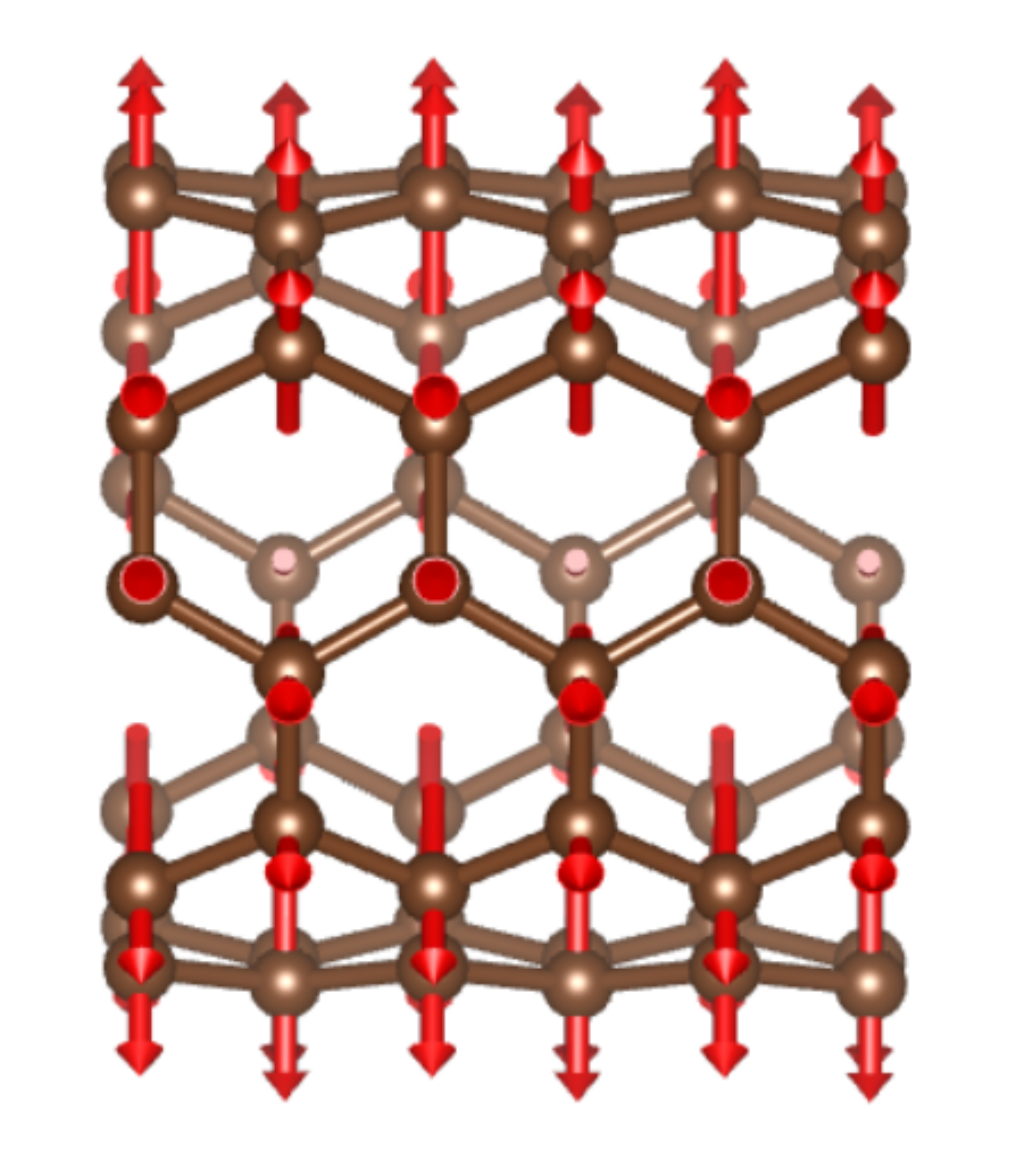}
\includegraphics[width=0.22\hsize, bb=0 0 295 313]{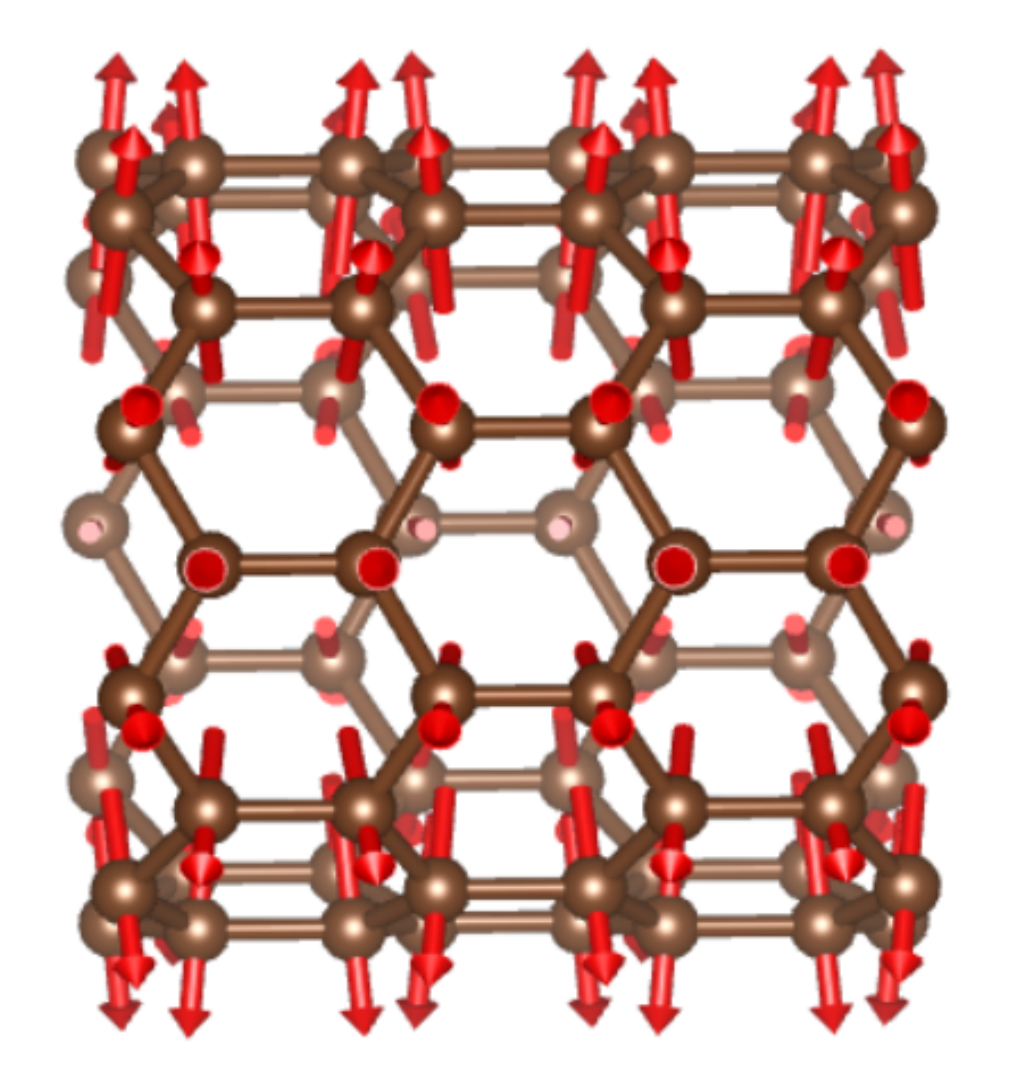}
\includegraphics[width=0.24\hsize, bb=0 0 316 331]{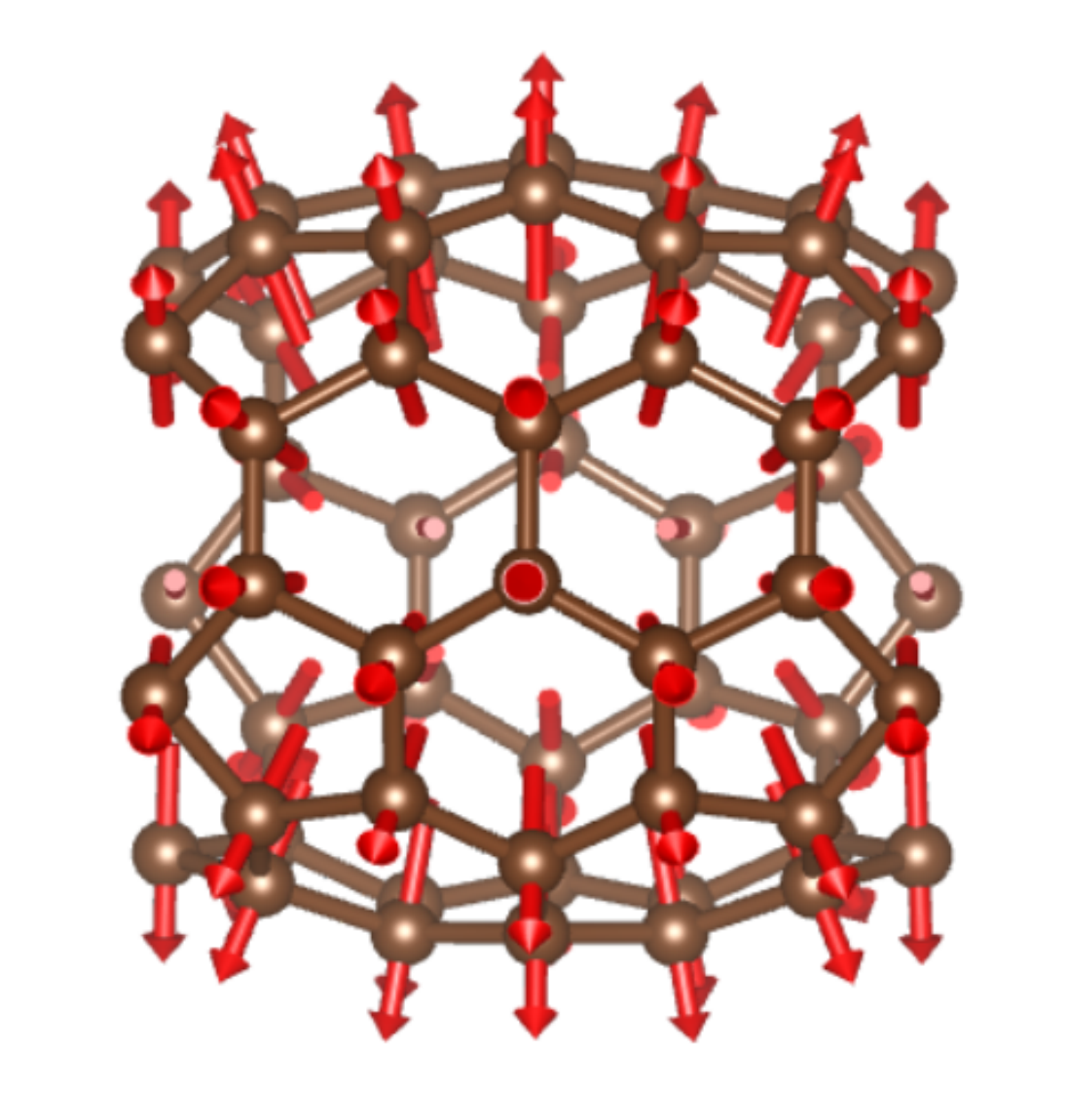}
\end{center}

\hskip 0.10\hsize (a)
\hskip 0.19\hsize (b)
\hskip 0.19\hsize (c)
\hskip 0.18\hsize (d)

\begin{center}
\includegraphics[width=0.25\hsize, bb=0 0 318 300]{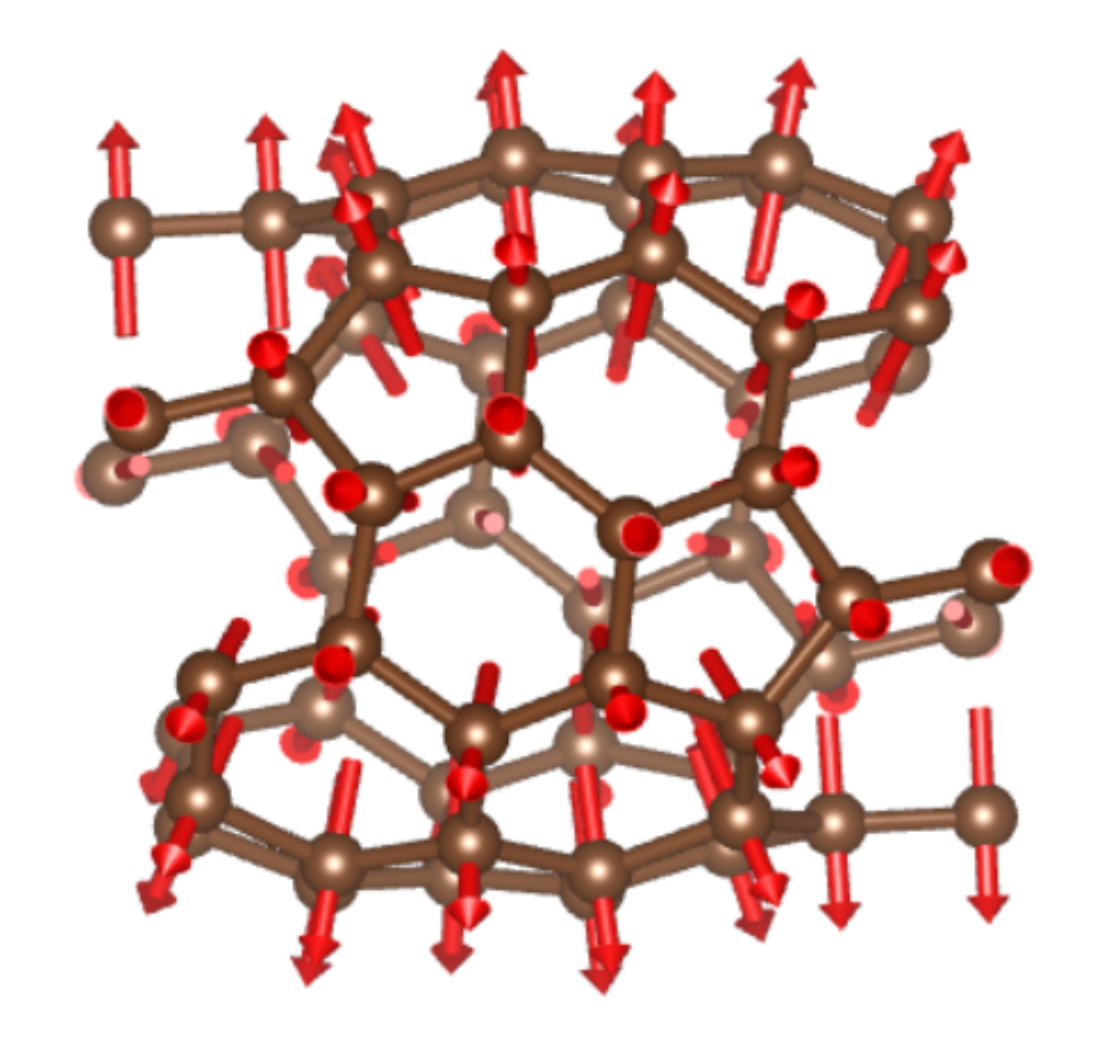}
\includegraphics[width=0.25\hsize, bb=0 0 322 297]{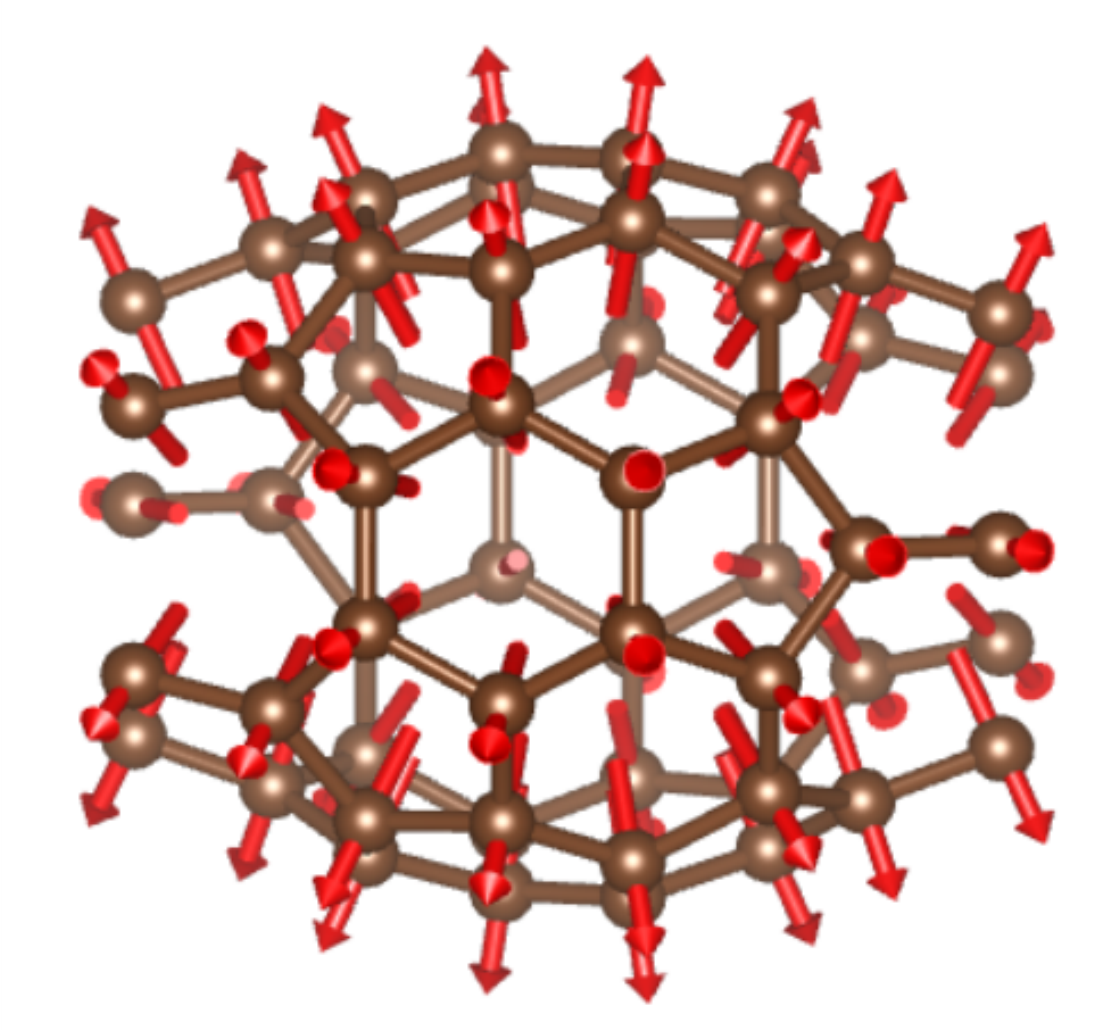}
\includegraphics[width=0.43\hsize, bb=0 0 392 204]{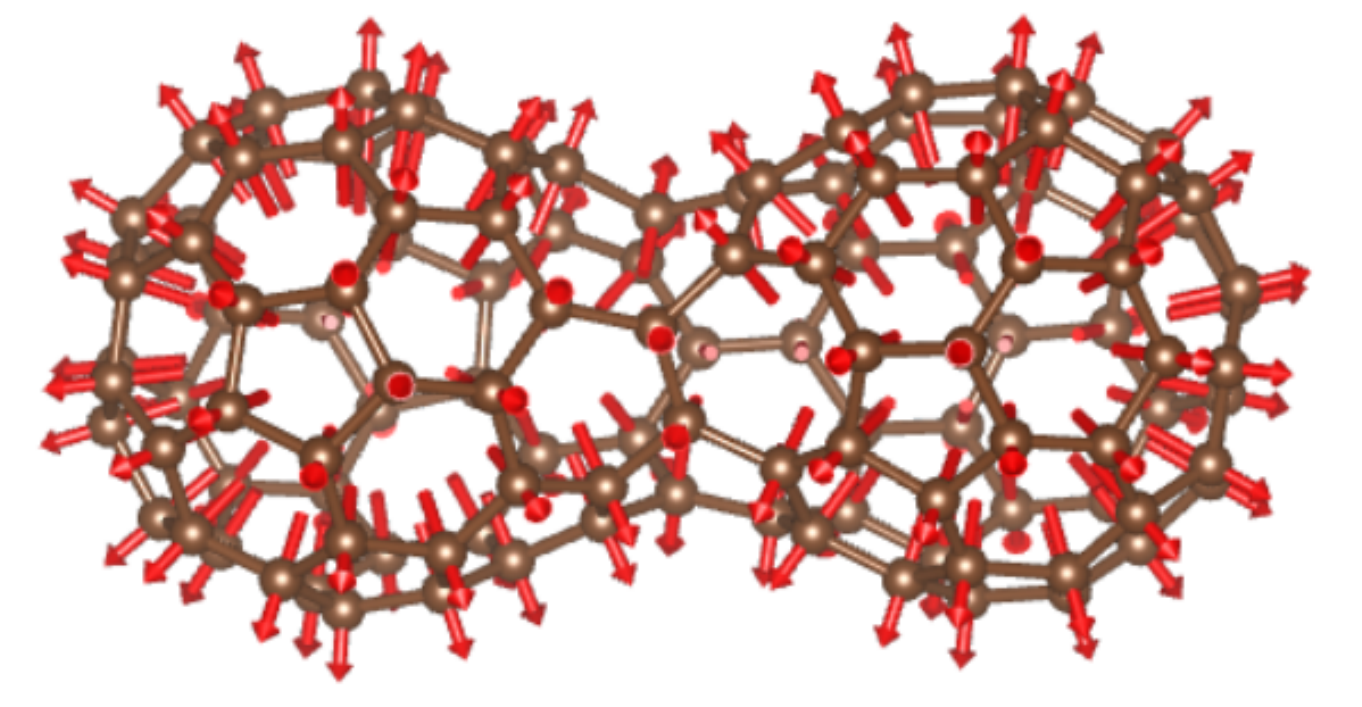}
\end{center}

\hskip 0.12\hsize (e)
\hskip 0.19\hsize (f)
\hskip 0.31\hsize (g)

\begin{center}

\caption{
The shape of nanocarbons by the first-principles calculations:
 C$_{60}$ (a),nanotubes 05-05 (b) and 09-00 (c),
the C$_{60}$-polymers, FP5N (d), FP4L (e), T3 (f), and C$_{60}$ dimer (C$_{120}$, P08) (g).
The arrows show the discrete normal vector at each carbon.
}\label{fg:Fullns}

\end{center}

\end{figure}

Since C$_{60}$ and carbon nanotubes (CNT) are expected to have the CDPC, we evaluate the discrete curvatures in the nanocarbons based on as illustrated in Figure \ref{fg:Fullns} (each arrow indicates a normal vector).
Their geometrical structures were determined by the first principles calculations. Here, the parameters are the same as those described above \cite{NOO}.
We examined the two types of the carbon nanotubes as shown in Figures \ref{fg:Fullns} (b) and (c).
Nanotube 05-05 consists of ten carbon atoms with one of the edges of the carbon hexagon parallel to the perimeter, whereas nanotube 09-00 has nine perimeter carbon atoms with one of the carbon hexagon parallel to the center axis.
Figures \ref{fg:Fullns} (d)-(f) show three types (FP5N, FP4L, and T3) of the peanut-shaped C$_{60}$ polymers, respectively. 
Figure \ref{fg:Fullns} (g) shows a peanut-shaped C$_{60}$ dimer, (C$_{120}$, P08) reported previously \cite{TON}.

\begin{table}[htb]
\centering
\caption{The discrete principal curvature of the nanocarbons:}\label{tbl:DPC_full}
  \begin{tabular}{c|cccccc}

Nanocarbons& $\overline{k_1}$& $\delta k_1$& $1/\overline{k_1}$& $\overline{k_2}$& $\delta k_2$& $1/\overline{k_2}$\\
 & $(\AA^{-1})$& $(\AA^{-1})$& $(\AA)$& $(\AA^{-1})$& $(\AA^{-1})$& $(\AA)$\\
\midrule
C$_{60}$ & 0.2875&  0.0004 & 3.4781& 0.2740&  0.0003& 3.6501\\
CNT 05-05 &0.2891& 0.0041& 3.4586&0.0000 &0.0000&---\\
CNT 09-00& 0.2784&0.0070&3.5923&0.0028 &0.0008&---\\
\hline
PSFP FP5N&0.3061&0.0224&3.2674&0.0344&0.1207&29.0469\\
PSFP FP4L&0.3075&0.0344&3.2518&0.0348&0.1263&28.7248\\
PSFP T3 &0.3216&0.0490&3.1092&0.0644&0.1675&15.5097\\
\hline
C$_{120}$ P08  &0.3018&0.0404&3.3132&0.1694&0.1592&5.9019\\
\bottomrule
  \end{tabular}
\end{table}

\begin{table}[htb]
\centering
\caption{The angle and bond length of the nanocarbons:}\label{tbl:theta_a}
  \begin{tabular}{c|cccc}
Nanocarbons&$\overline{\theta}$& $\delta \theta$& $\overline{a}$& $\delta a$\\
    & (rad) & (rad) & $(\AA)$ & $(\AA)$ \\
\midrule
C$_{60}$ &0.3333& 0.0124& 1.4349& 0.0254\\
CNT 05-05 &0.3333&0.0010&1.4288&0.0006\\
CNT 09-00& 0.3333& 0.0027& 1.4288& 0.0059\\
\hline
PSFP FP5N&0.3333&0.0187&1.4334&0.0279\\
PSFP FP4L&0.3333&0.0181&1.4342&0.0285\\
PSFP T3 &0.3333&0.0151&1.4360&0.0375\\
\hline
C$_{120}$ P08  &0.3333&0.0142&1.4356&0.0286\\
\bottomrule
  \end{tabular}

\end{table}

The discrete principal curvatures $k_1$ and $k_2$ of these nanocarbons are summarized in Table \ref{tbl:DPC_full}; $\overline{k_a}$, $1/\overline{k_a}$, and $\delta k_a$ denote the average of $k_a$, the curvature radius, the standard deviation against the carbon atoms, respectively.
In order to confirm the calculated results of Table 2, we examined the average and the standard deviation of the angle $\theta$ and the bond length $a$ for each bond as shown in Table \ref{tbl:theta_a}.
Since they show the constant length and constant angle, the calculated results are reasonable.

Accordingly, Table \ref{tbl:DPC_full} shows the following interesting results:

\begin{enumerate}

\item 
Since all of the standard deviation of the first principal curvature of $\delta k_1$ is sufficiently smaller than $k_1$ itself, they are classified to the pCDPC surfaces.

\item 
Since all of the average of $k_1$ is similar, and  its inverse, or the curvature radius is nearly equal to 3.47 $\AA$  of C$_{60}$, the property does not depend on the computational parameters of the first-principles calculations. Namely, FP5N, FP4L, and T3 have the same property.

\item 
C$_{60}$ and CNTs have the CDPC $k_a$ $(a=1,2)$ up to computational deviations or discrete nature. [As mentioned in \cite{KNO}, non-trivial center axisoid appears for the CNTs.]

\end{enumerate}

\begin{figure}[H]
\begin{center}
\includegraphics[width=0.35\hsize, bb=0 0 814 434]{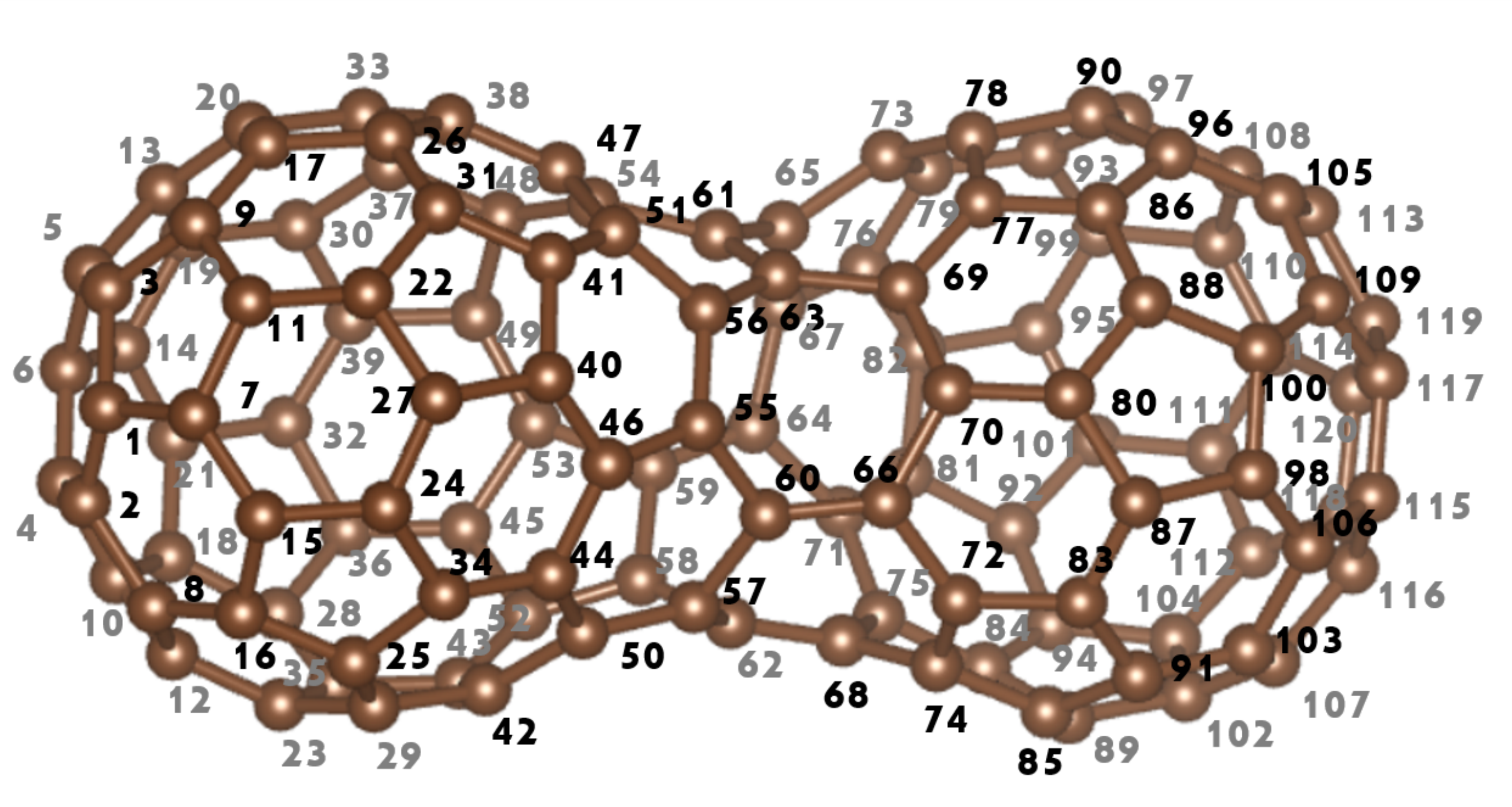}

 (a)

\includegraphics[width=0.50\hsize, bb=0 0 626 323]{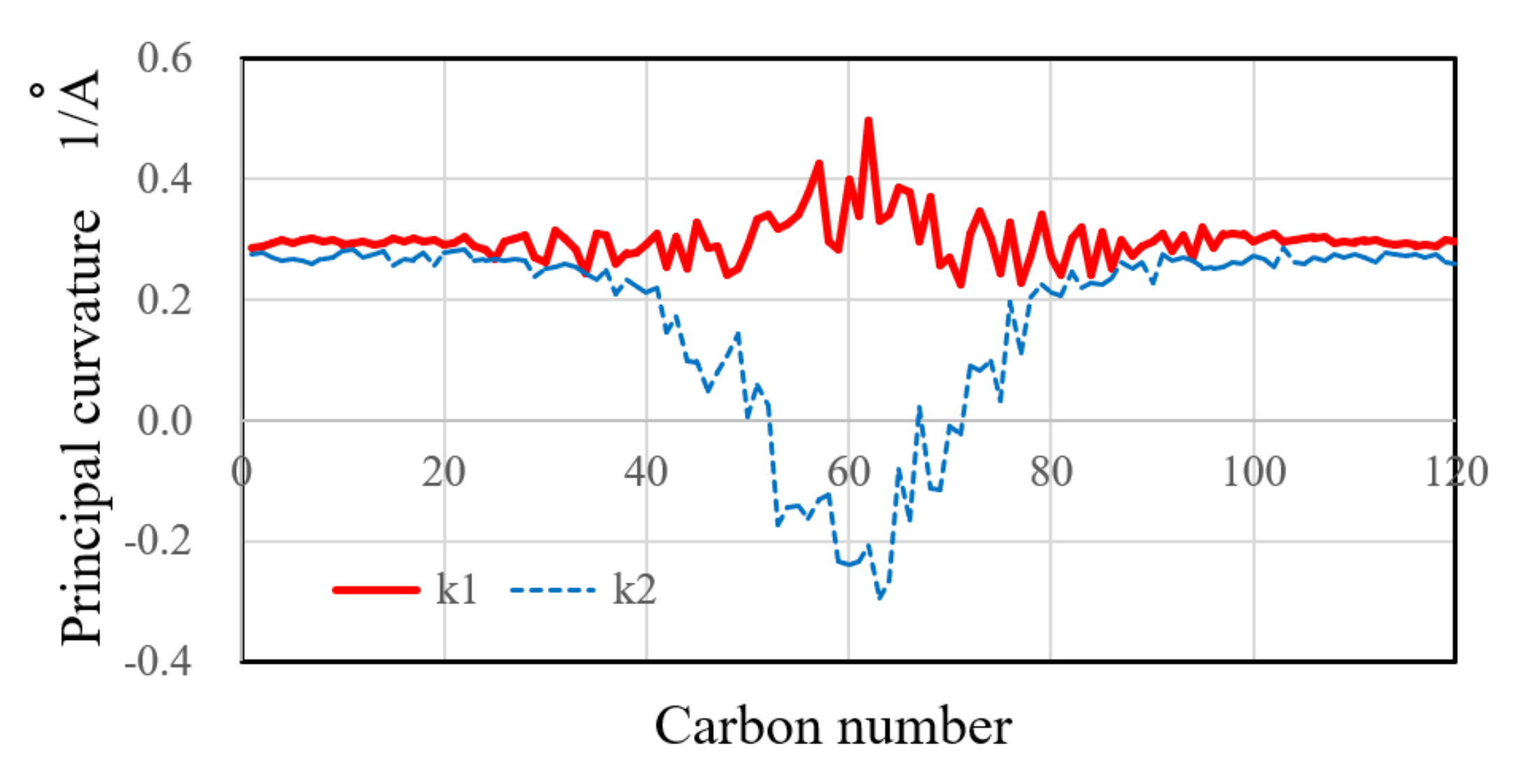}

 (b)

\includegraphics[width=0.50\hsize, bb=0 0 622 344]{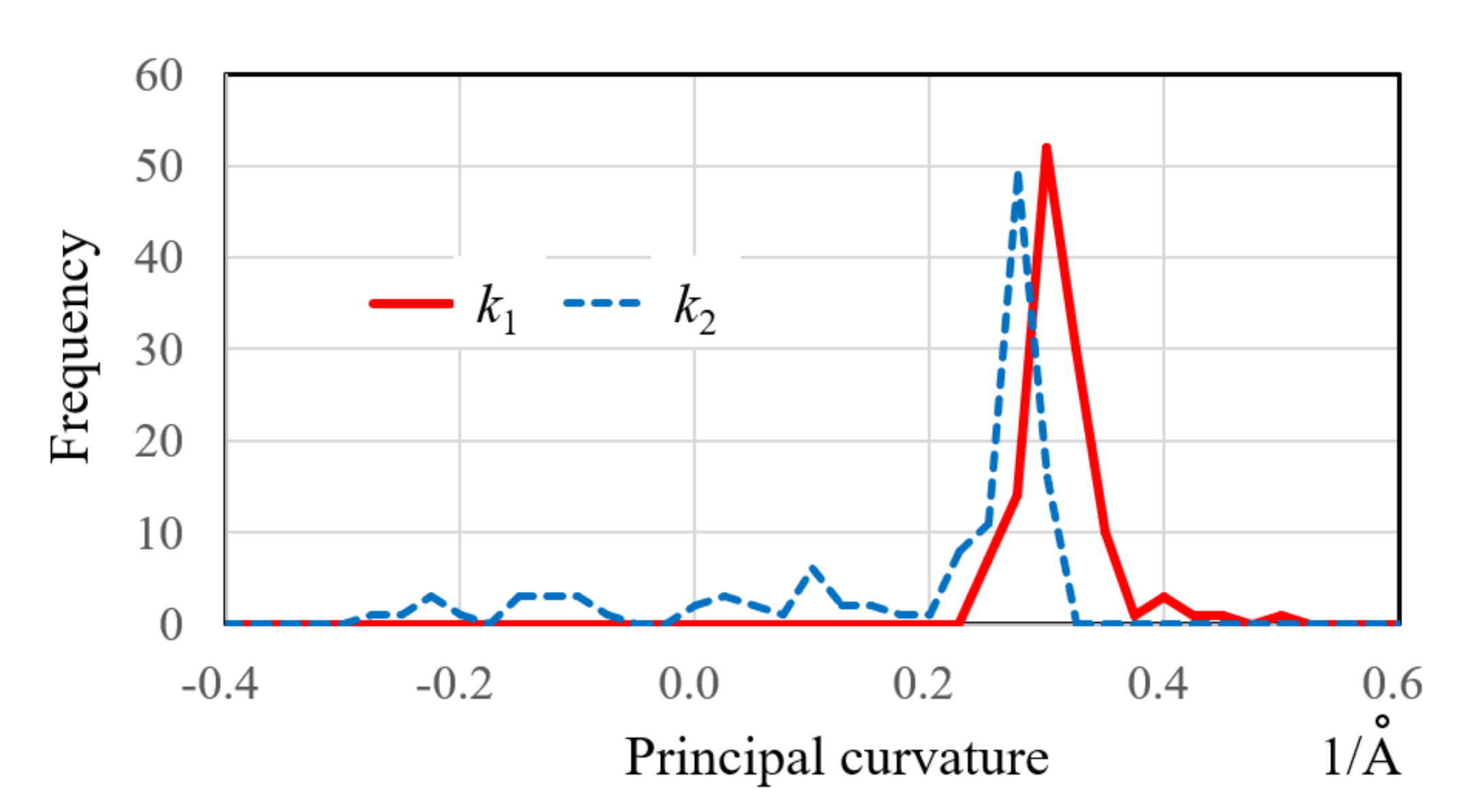}

(c)

\caption{
The discrete principal curvature distribution of C$_{60}$ dimer, (C$_{120}$, P08):
(a) shows the numbering of the carbon atoms in C$_{60}$ dimer, (C$_{120}$, P08), (b) the value of the discrete principal curvatures $k_a$ $(a=1,2)$ at each carbon atom, and (c) the frequency.
}\label{fg:P08_dist}

\end{center}
\end{figure}

In case of the C$_{60}$ dimer, (C$_{120}$, P08), the deviation $\delta k_1$ of P08 is not small when compared to the others and its distribution as shown in Figure \ref{fg:P08_dist}, which shows that it may be also referred to the pCDPC surface.
Since the CDPC is a special case of pCDPC, it turns out that the nanocarbons we considered have the pCDPC.
The geometrical property is fascinating, which has never been mentioned before.

\section{Discussion}\label{sec:S5}

We geometrically analyzed the configuration data of the carbon atoms for the nanocarbons (see Figure \ref{fg:Fullns}) and investigated their geometrical properties, especially discrete principal curvature distributions. 
It is found that all the nanocarbon we investigated have almost constant discrete principal curvature distributions, which are classified to the pCDPC surfaces;
Such geometrical property has never been mentioned before. 
It is due to a nature of the nanocarbons. 
Since these nanocarbons except two CNTs can be produced from C$_{60}$ via the general Stone-Wales transformation \cite{NOO}, the nature of the nanocarbons is considered to originate from the pCDPC properties.

\begin{figure}[H]
\begin{center}
\includegraphics[width=0.35\hsize, bb=0 0 576 401]{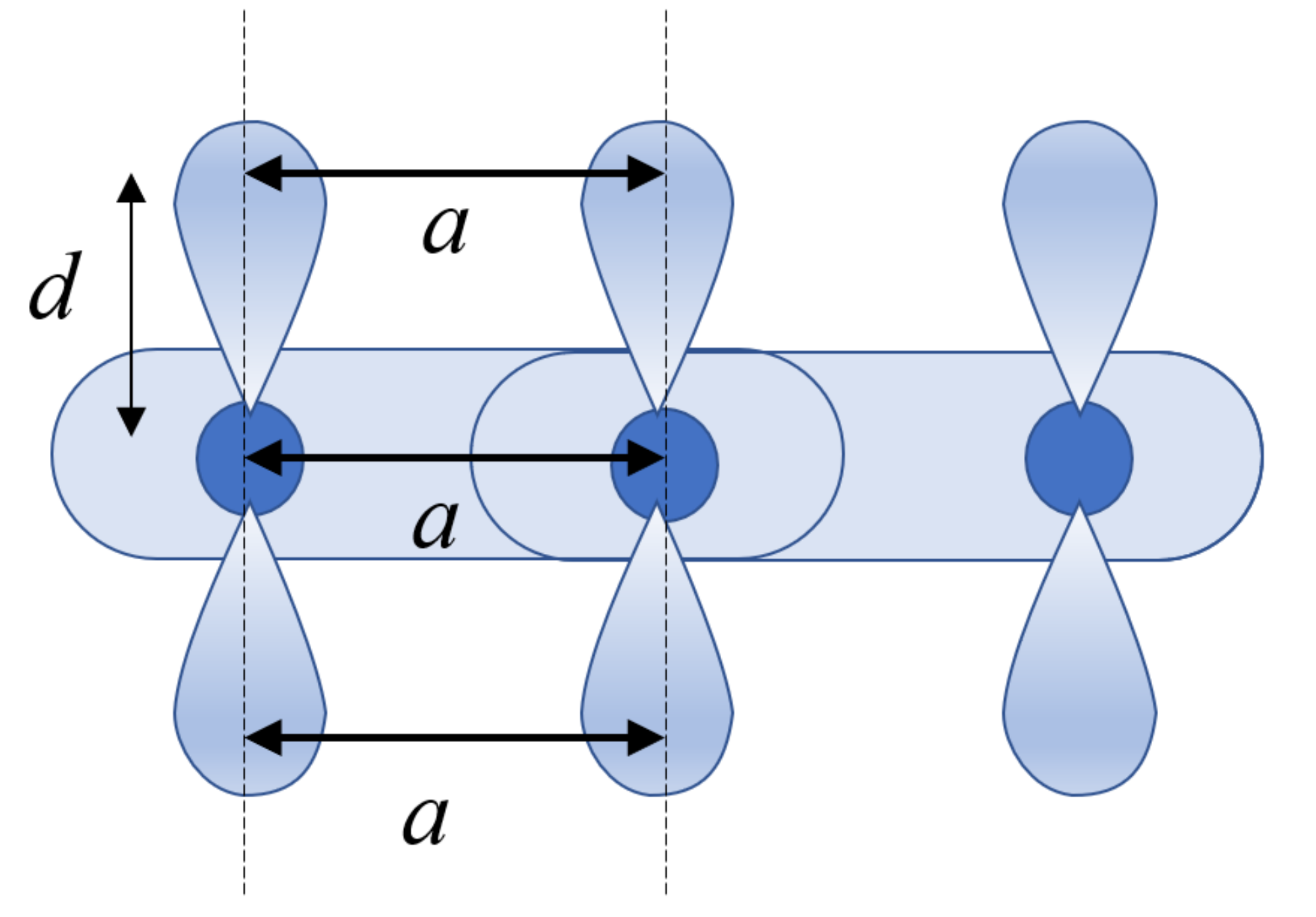}
\includegraphics[width=0.35\hsize, bb=0 0 646 600]{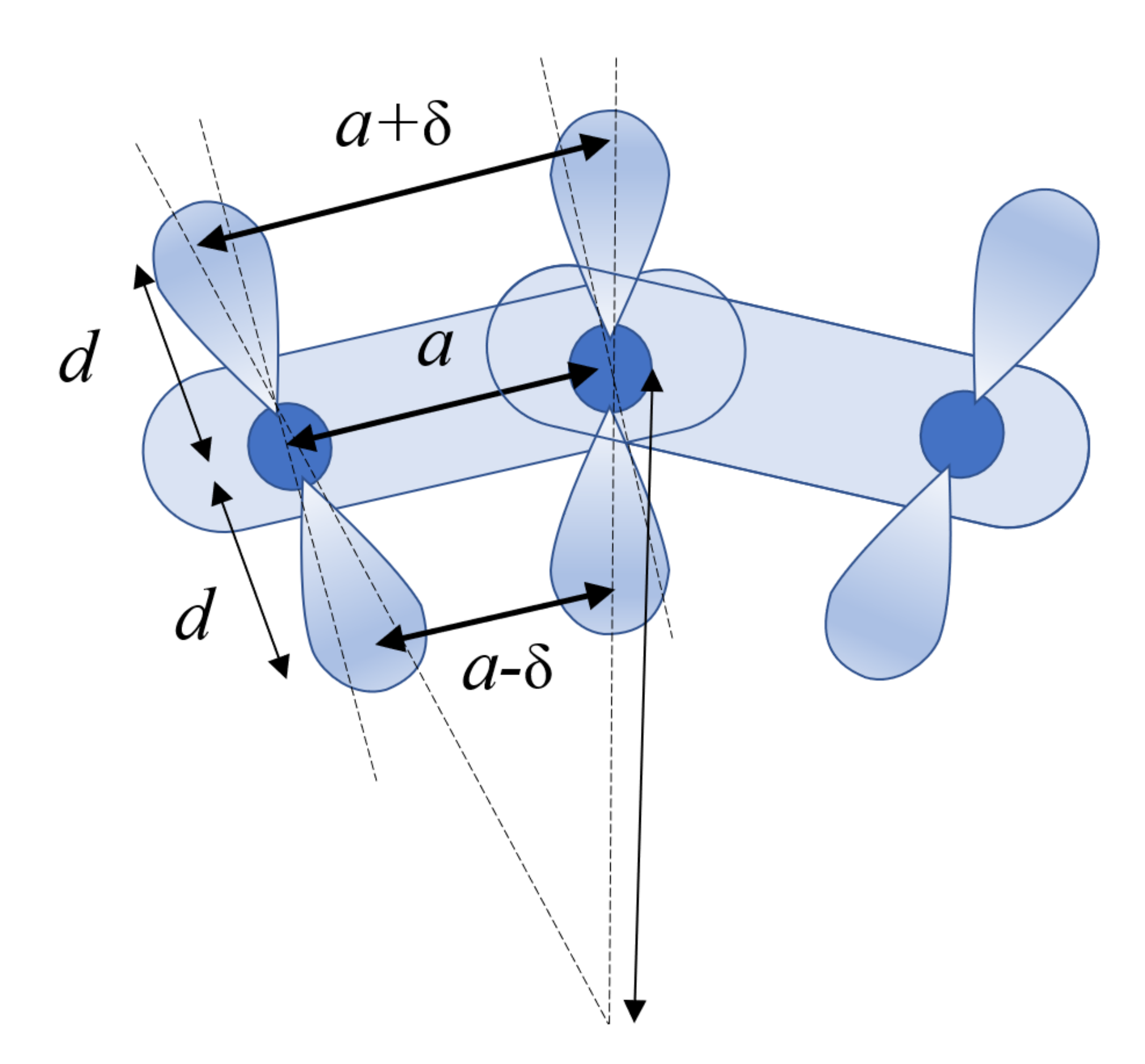}

\hskip 0.00\hsize (a)
\hskip 0.30\hsize (b)

\caption{
The curvature effect on the C-$2p$ atomic orbitals in the tight binding picture:
(a) the flat cases, (b) the extra overlap integrals due to the discrete curvature.}\label{fg:Wfunc}

\end{center}

\end{figure}

It is noted that the elliptic ($k_1 k_2 > 0$) region of the C$_{60}$-polymer and C$_{60}$ dimer (P08) take over the part of C$_{60}$, and the other parts also preserve the connection as discrete graphs of C$_{60}$.
On the other hand, under the tight binding picture, the wave functions of each C-$2p$ atomic orbital are affected by the discrete curvature illustrated in Figure \ref{fg:Wfunc}.
The discrete curvature causes the extra overlap integral between adjacent C-$2p$ atomic orbitals, and the total energy correspondingly increases.
The discrete curvature expresses the curved carbon networks better than curvature in the continuum picture.

Some restriction of the graph network of C$_{60}$ determines the configuration of the carbon atoms so that the the overlap integral is globally minimized.
The directions of the principal curvatures, $k_1$ and $k_2$, are orthogonal to each other, even for the discrete picture \cite{KMO}. 
Then, one direction should be chosen because the smaller radius of curvature (or the larger (first) principal curvature, $|k_1| \ge |k_2|$) affects the overlap between adjacent the C-$2p$ atomic orbitals more than the other geometrical parameters, and correspondingly the total energy increases. 
This indicates that the first discrete principal curvatures should be constant; If a part deviates from the others, the overlap integral has deviation, and the energy density has distribution (not homogeneous).   
Thus, the first discrete principal curvature tends to be homogeneity.
In other words, the nature of the nanocarbon systems originating from C$_{60}$ makes the first discrete principal curvature constant rather than the second one.
Therefore, it can be predicted that the total energy would have something to do with the pre-constant first discrete principal curvature property for the C$_{60}$-polymers and C$_{60}$-dimer, (C$_{120}$, P08).

\begin{figure}[H]
\begin{center}
\includegraphics[width=0.24\hsize,bb= 0 0 1224 790]{FP5N.pdf}
\includegraphics[width=0.24\hsize,bb= 0 0 1224 790]{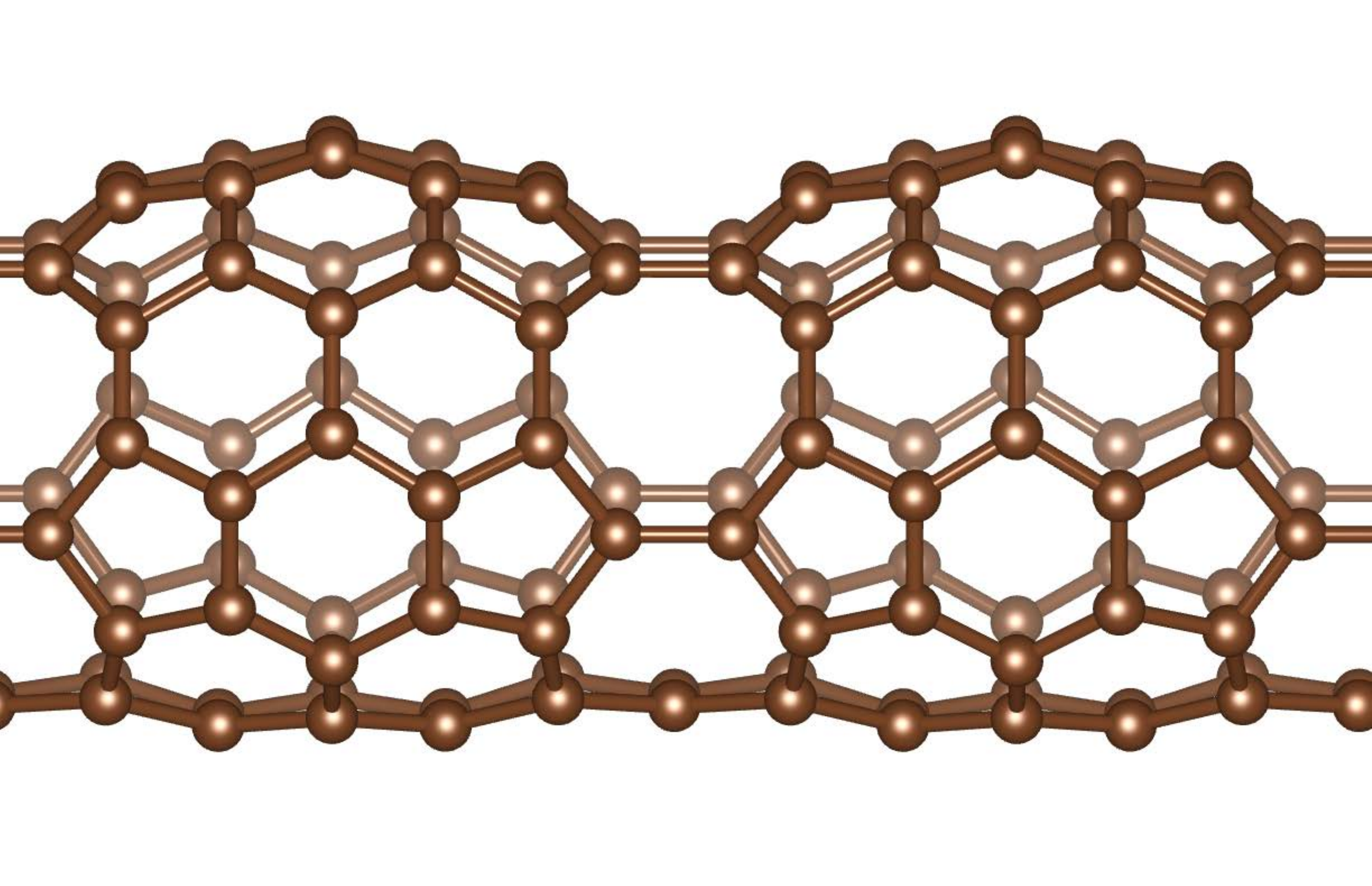}
\includegraphics[width=0.24\hsize,bb= 0 0 1224 790]{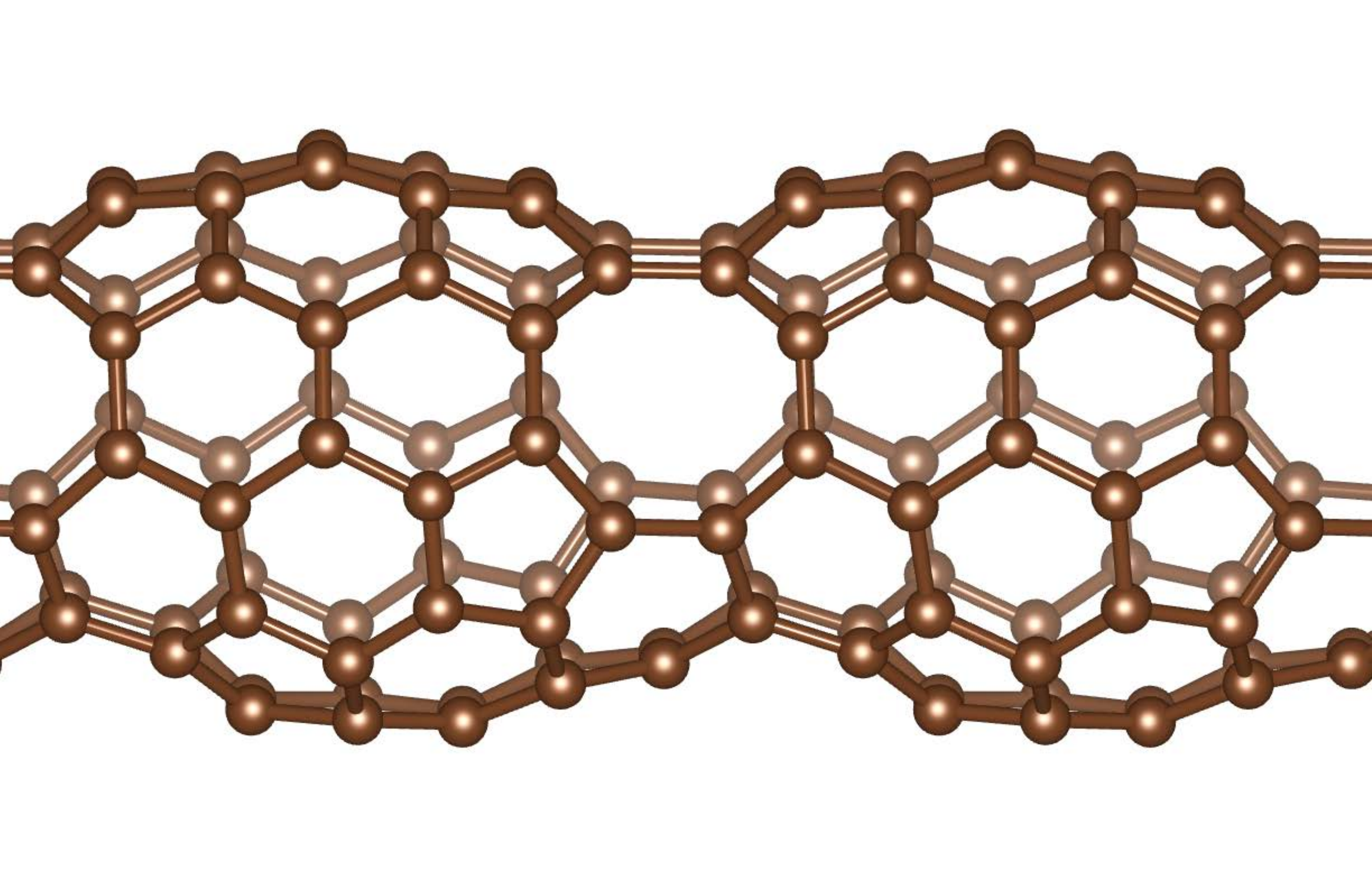}
\includegraphics[width=0.24\hsize,bb= 0 0 1224 790]{FP4L.pdf}

 (a)
\hskip 0.18\hsize (b)
\hskip 0.18\hsize (c)
\hskip 0.18\hsize (d)

\includegraphics[width=0.24\hsize,bb= 0 0 1224 790]{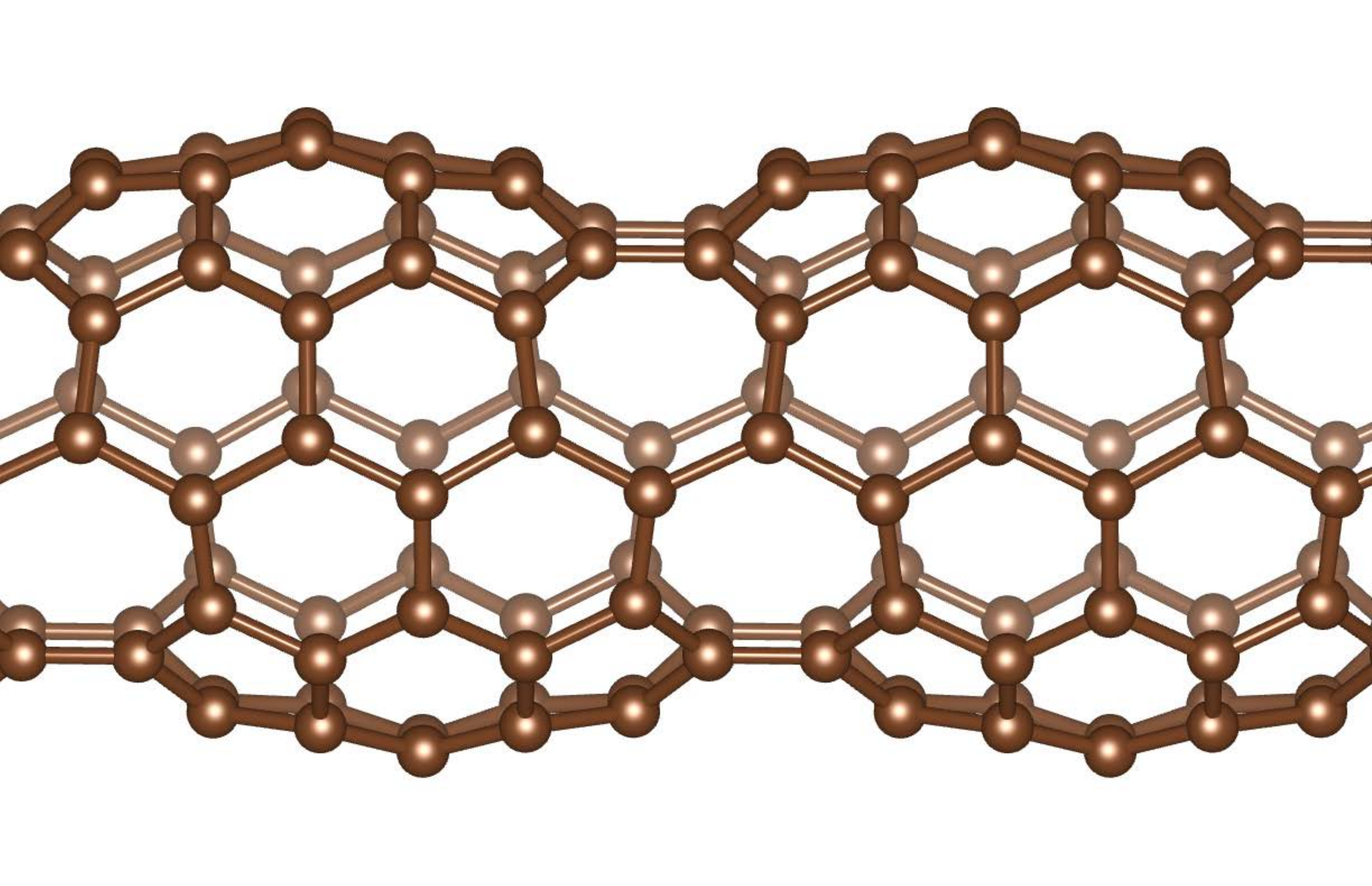}
\includegraphics[width=0.24\hsize,bb= 0 0 1224 790]{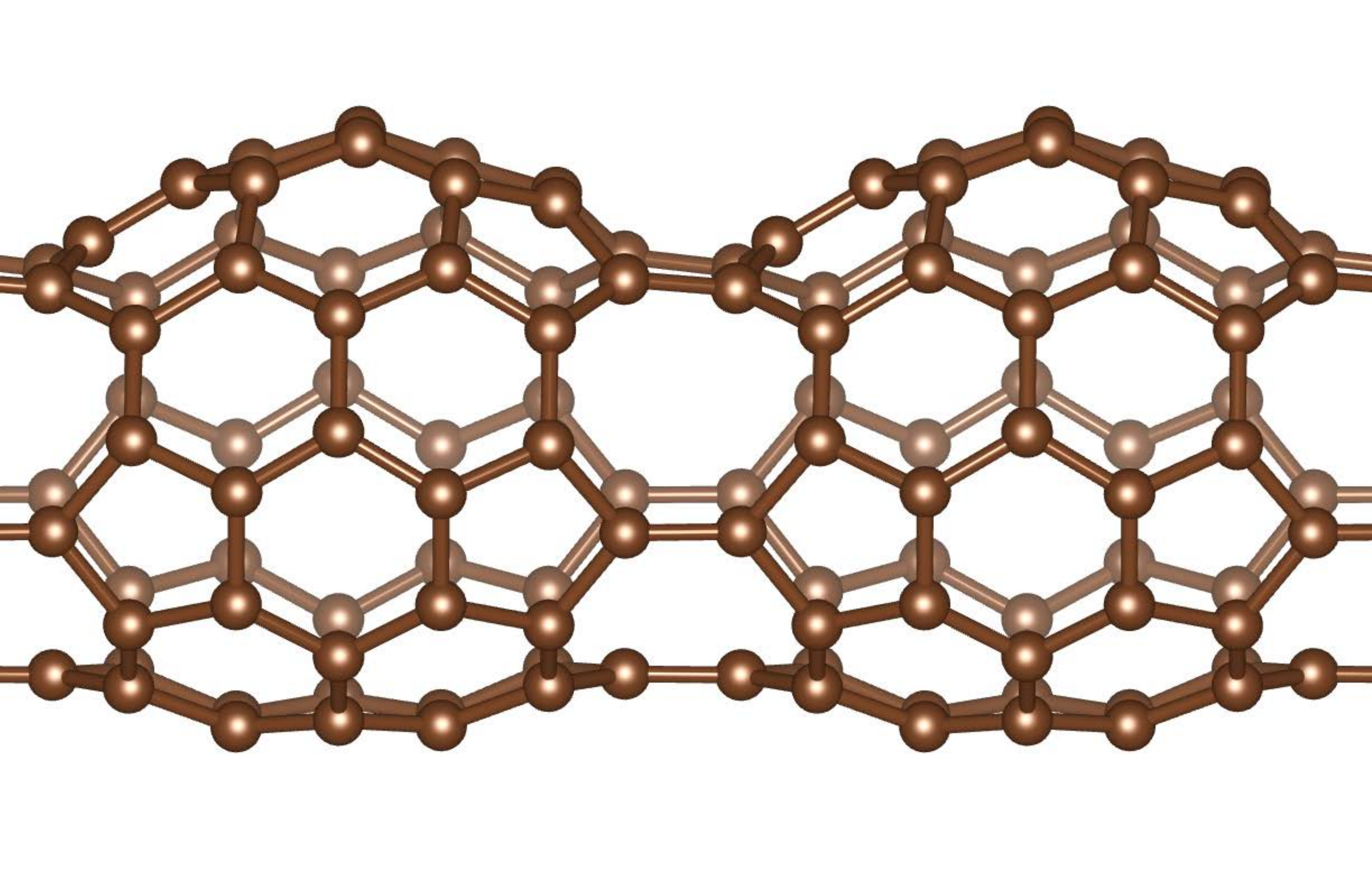}
\includegraphics[width=0.24\hsize,bb= 0 0 1224 790]{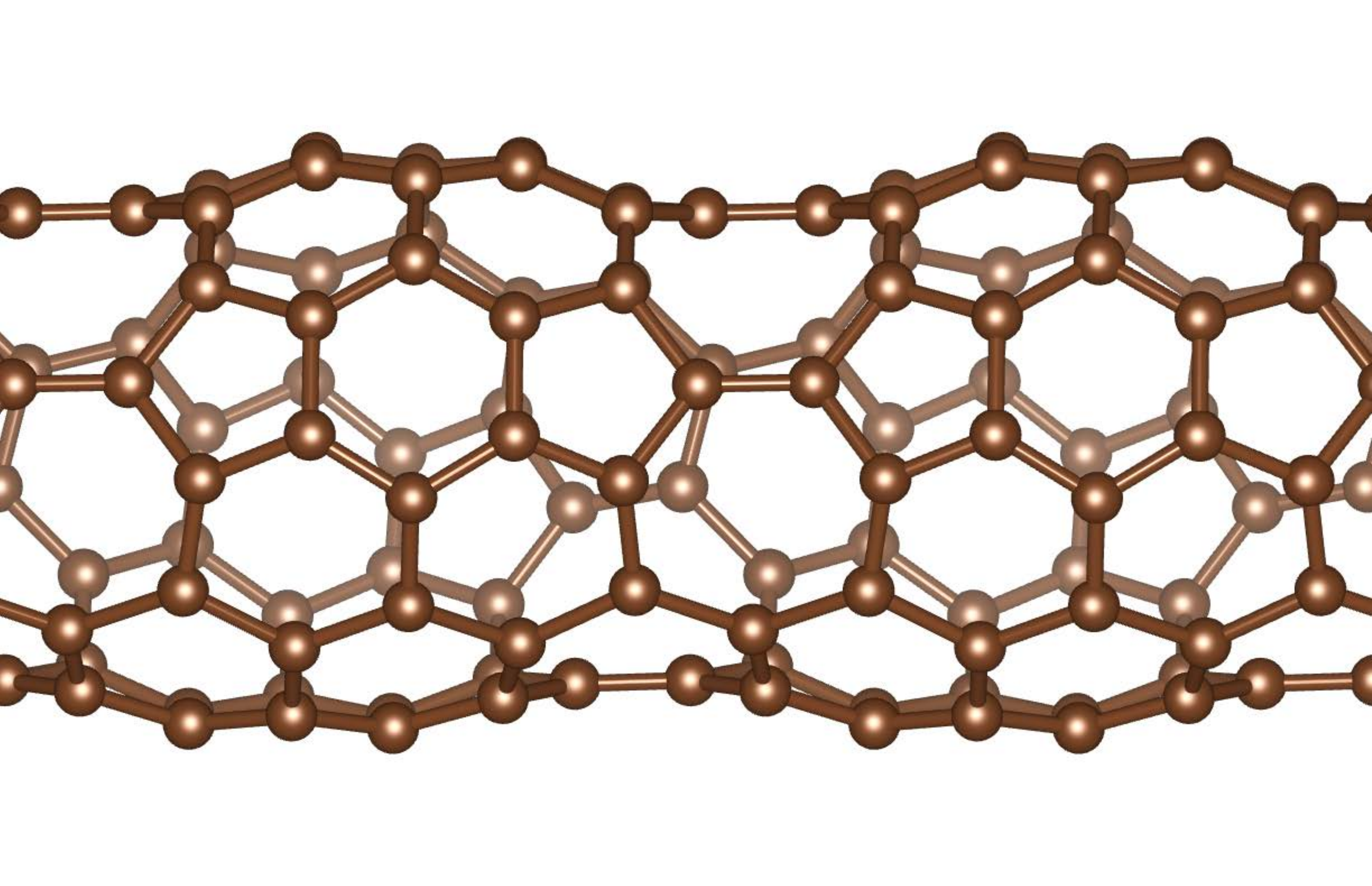}
\includegraphics[width=0.24\hsize,bb= 0 0 1224 790]{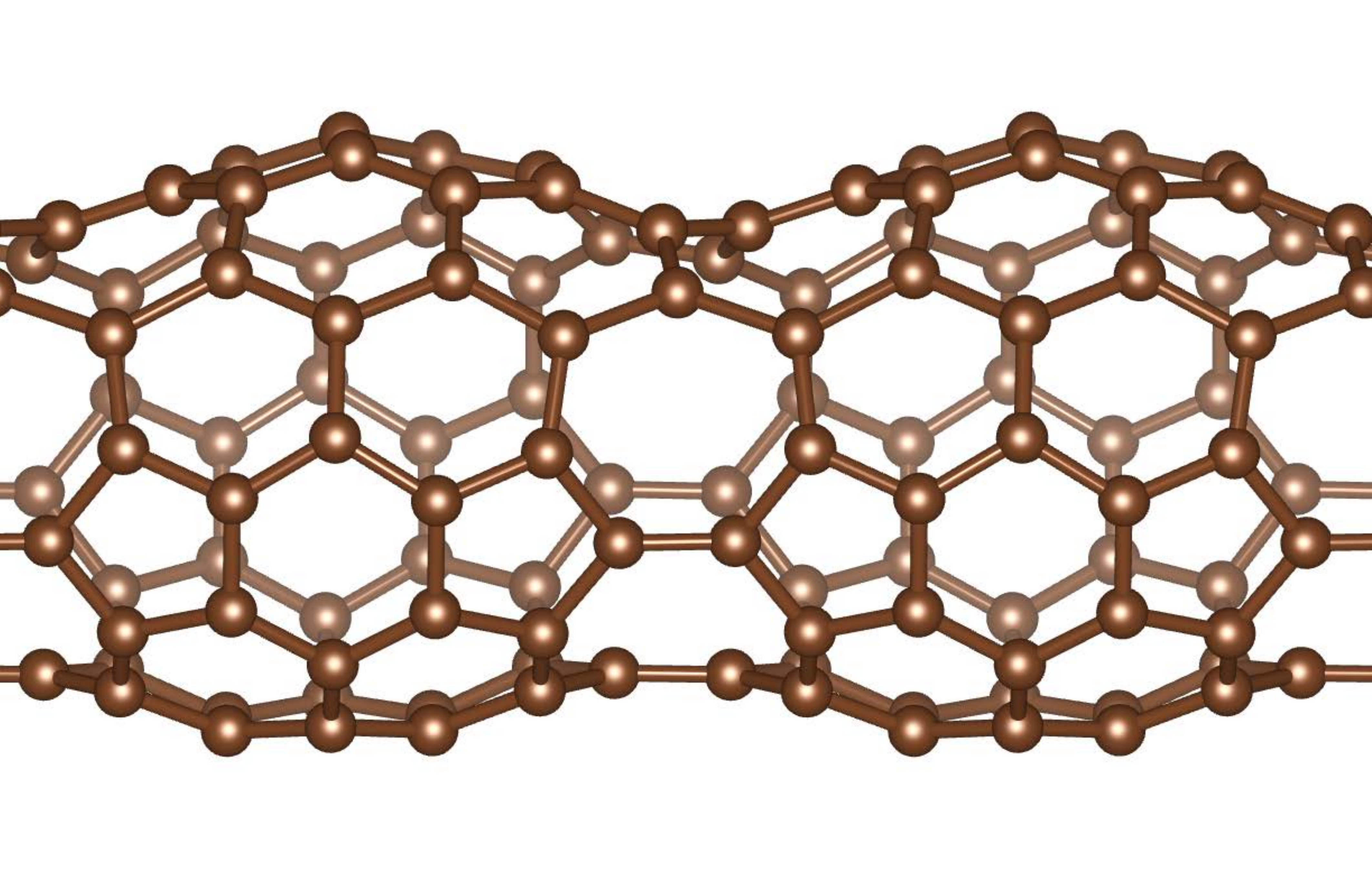}

 (e)
\hskip 0.18\hsize (f)
\hskip 0.18\hsize (g)
\hskip 0.18\hsize (h)

\includegraphics[width=0.24\hsize,bb= 0 0 1224 790]{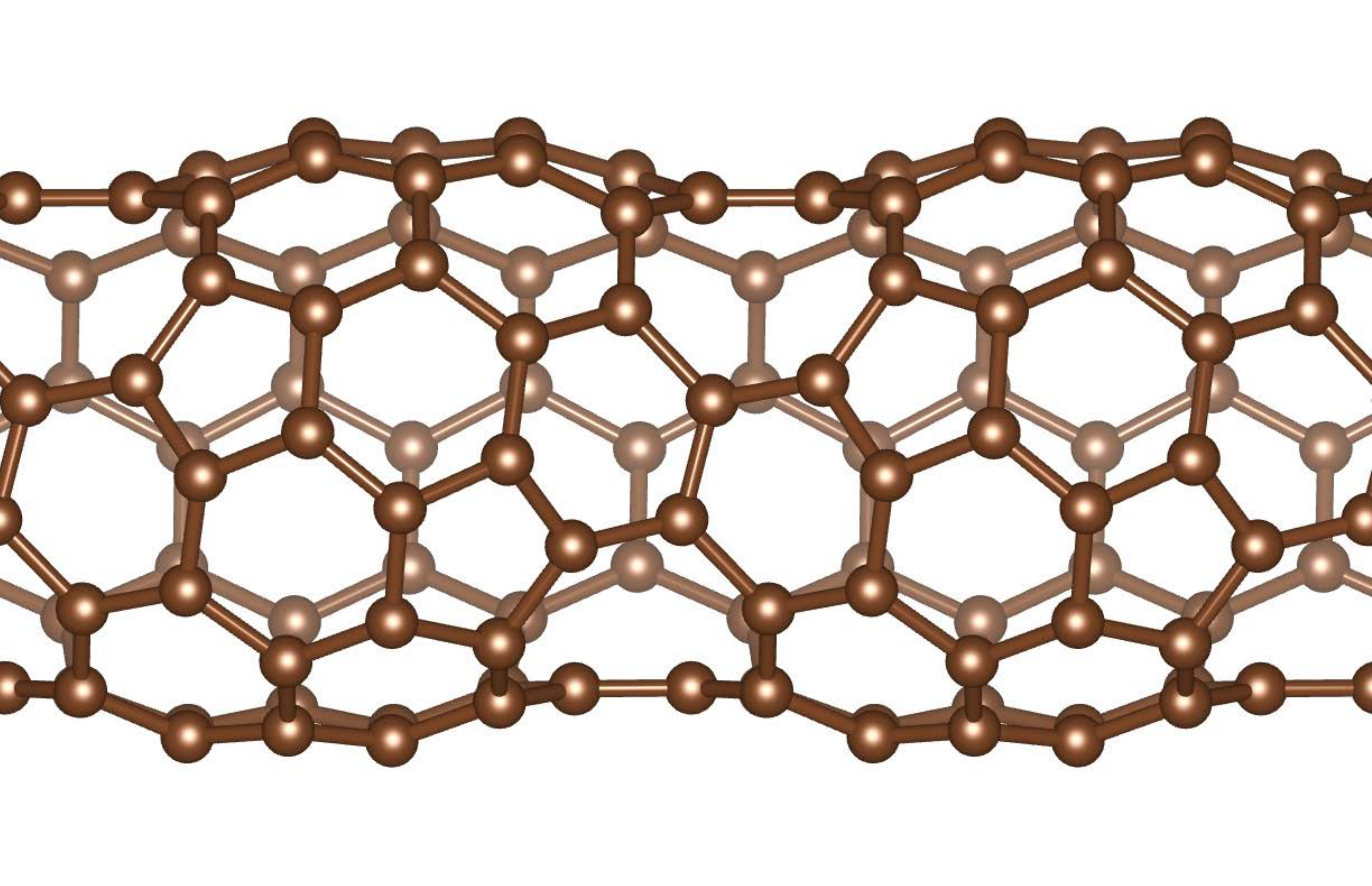}
\includegraphics[width=0.24\hsize,bb= 0 0 1224 790]{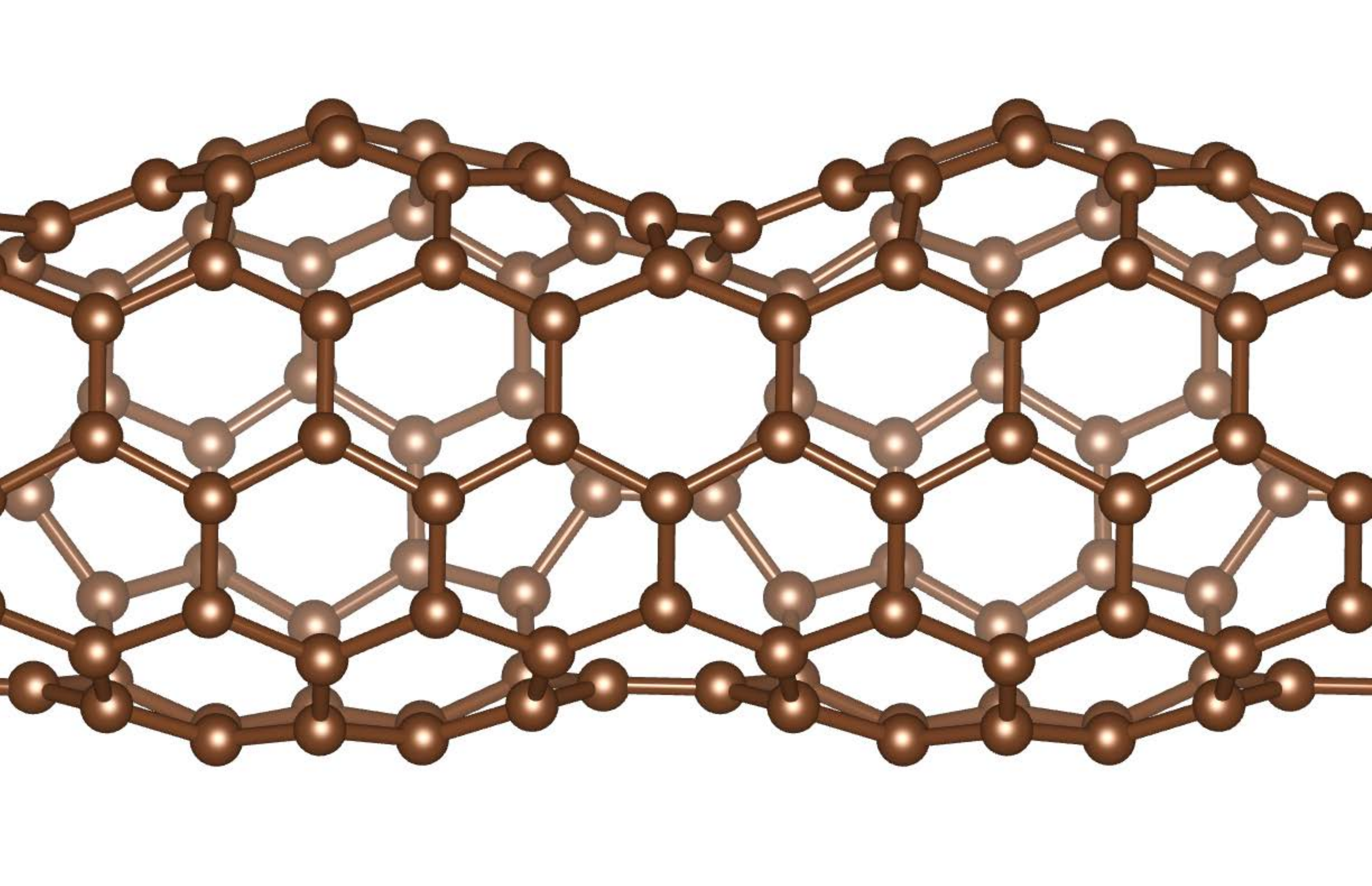}
\includegraphics[width=0.24\hsize,bb= 0 0 1224 790]{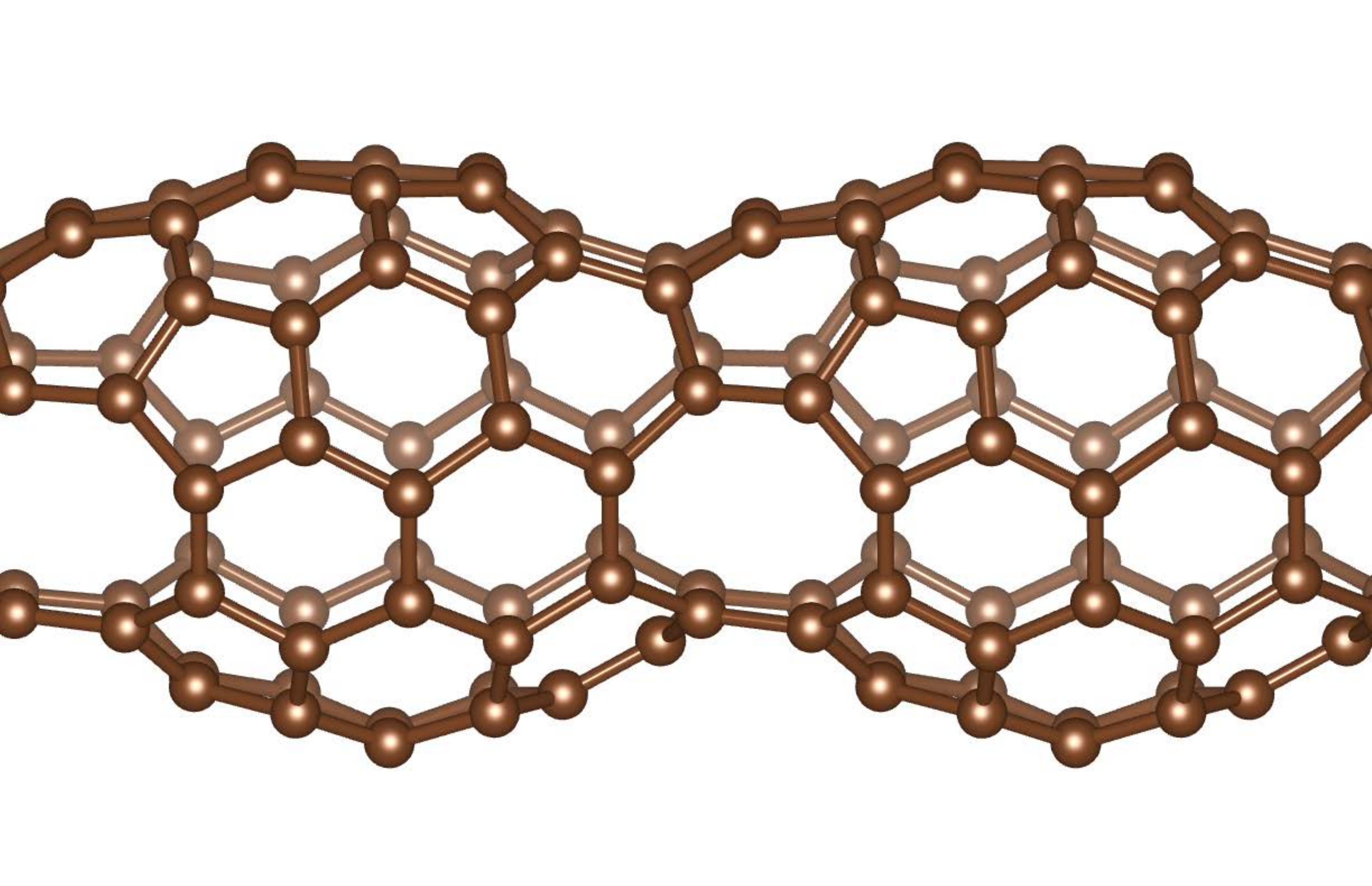}
\includegraphics[width=0.24\hsize,bb= 0 0 1224 790]{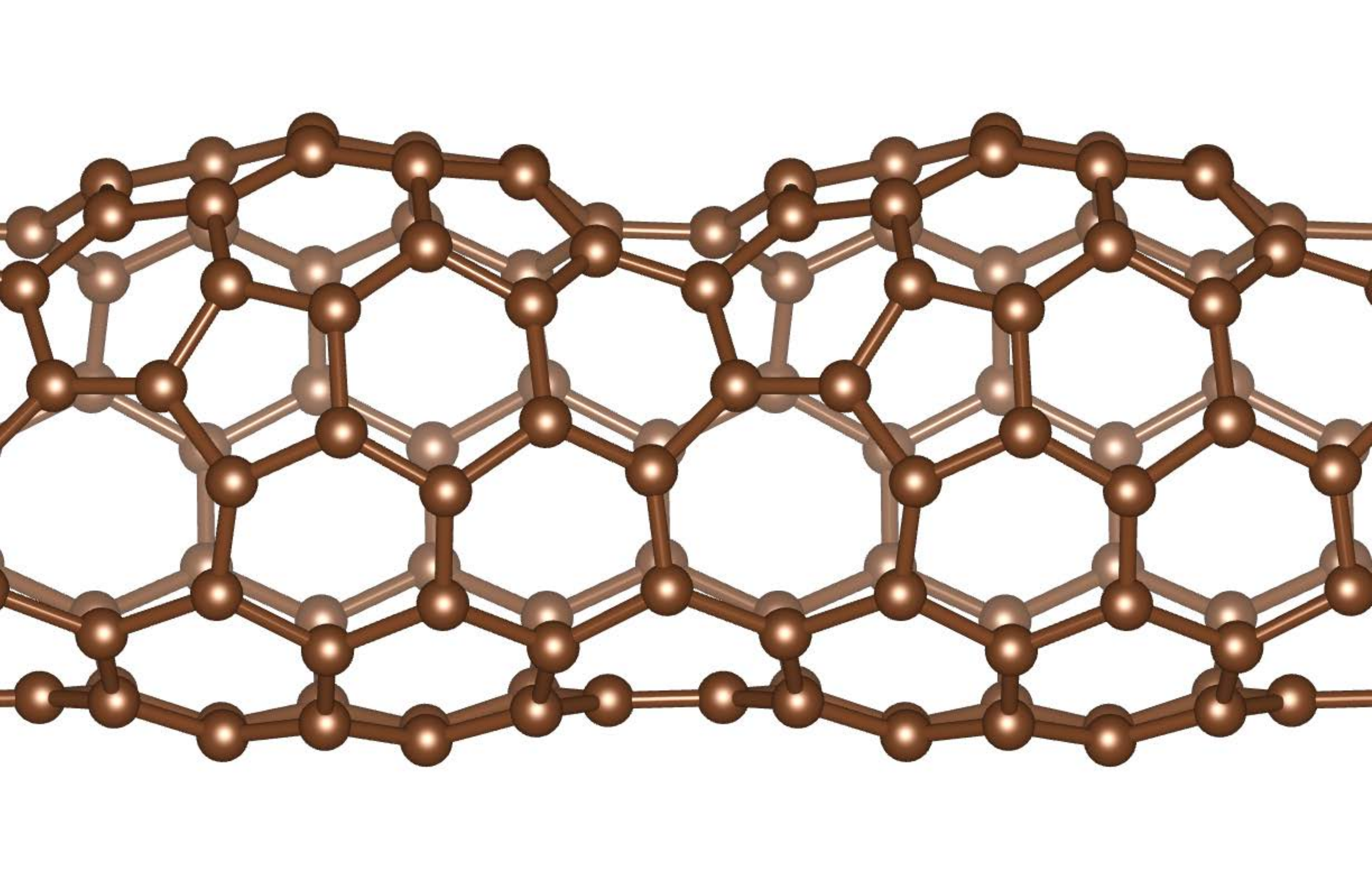}

 (i)
\hskip 0.18\hsize (j)
\hskip 0.18\hsize (k)
\hskip 0.18\hsize (l)

\includegraphics[width=0.24\hsize,bb= 0 0 1224 790]{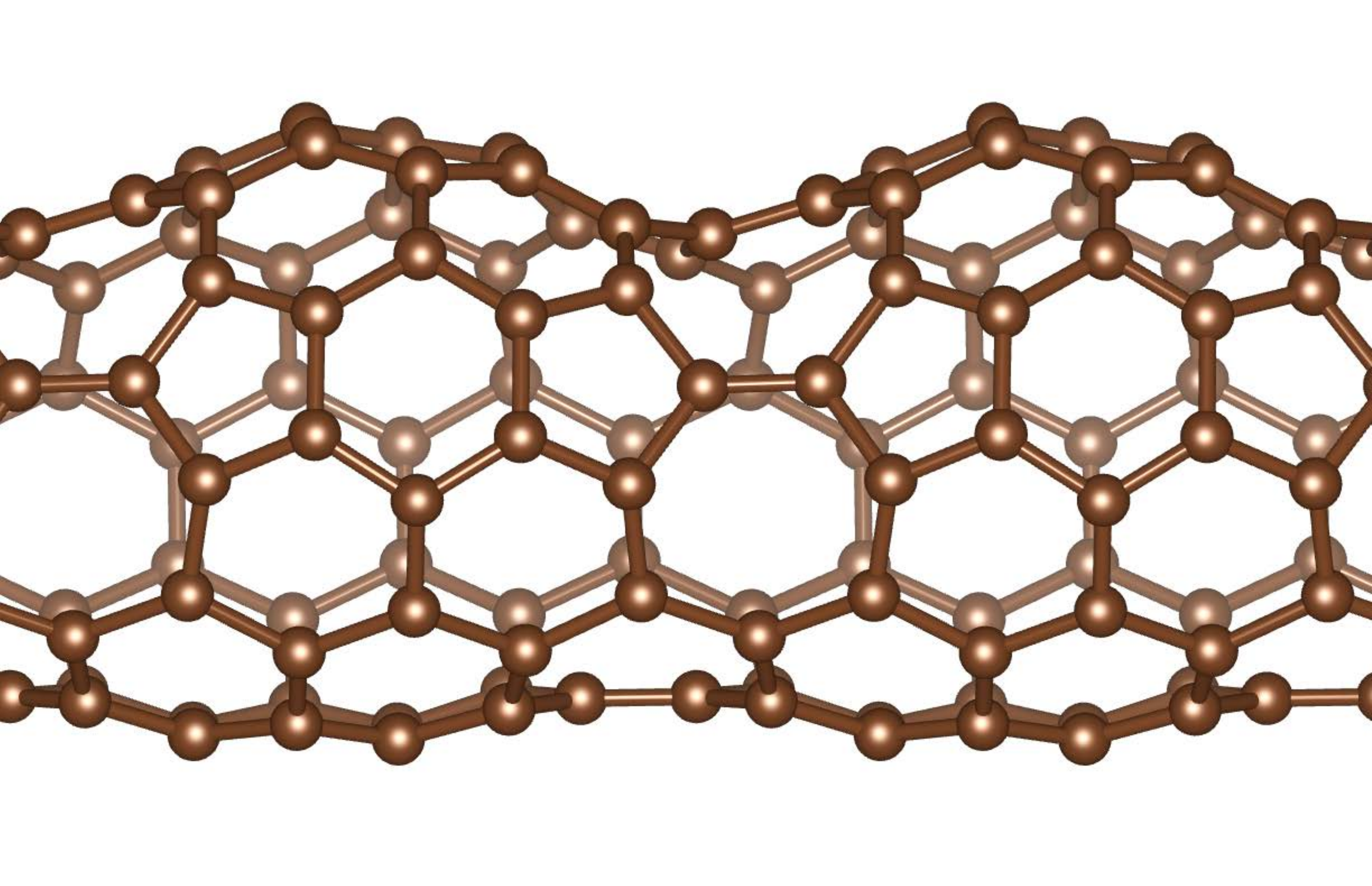}
\includegraphics[width=0.24\hsize,bb= 0 0 1224 790]{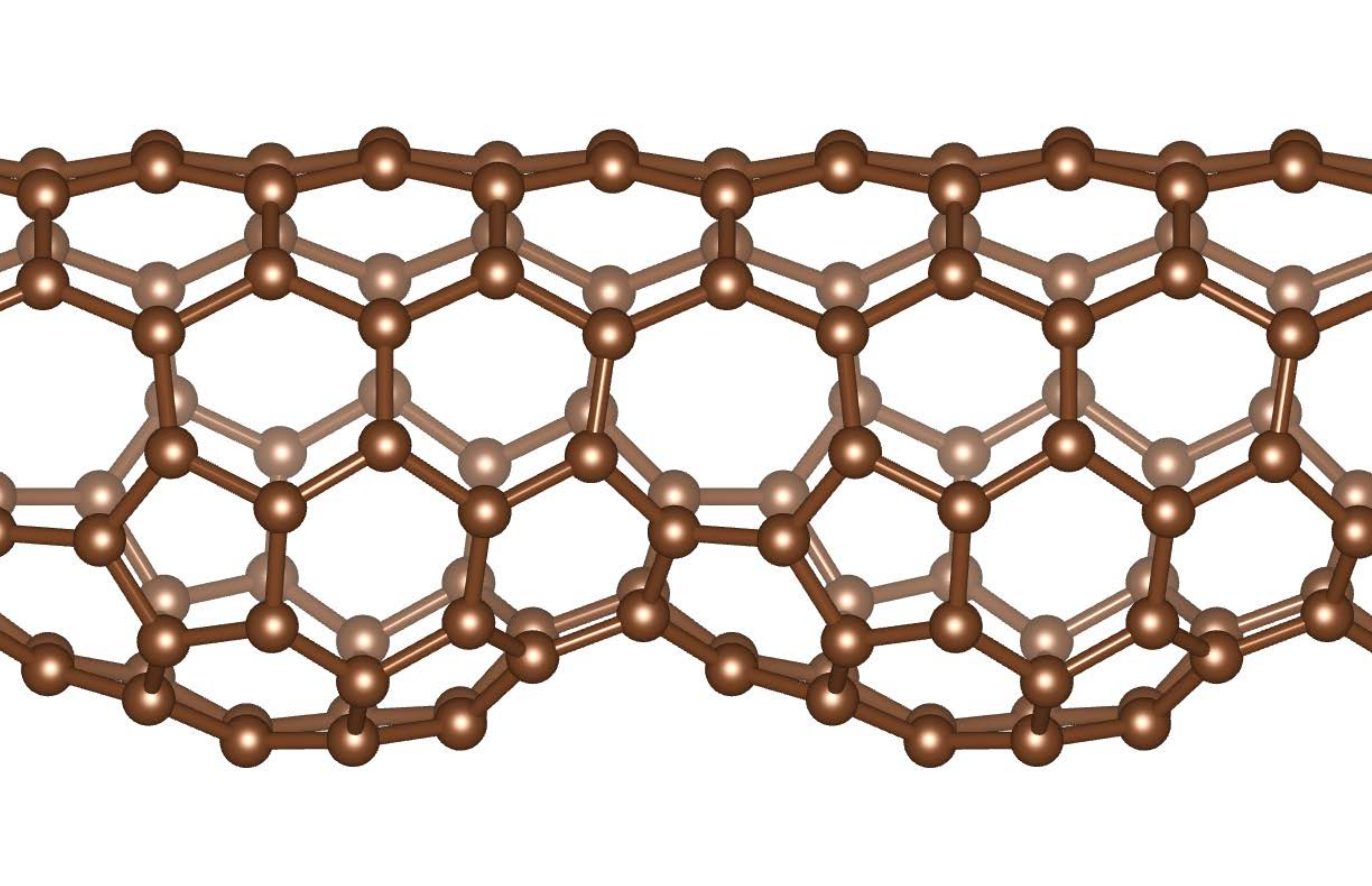}
\includegraphics[width=0.24\hsize,bb= 0 0 1224 790]{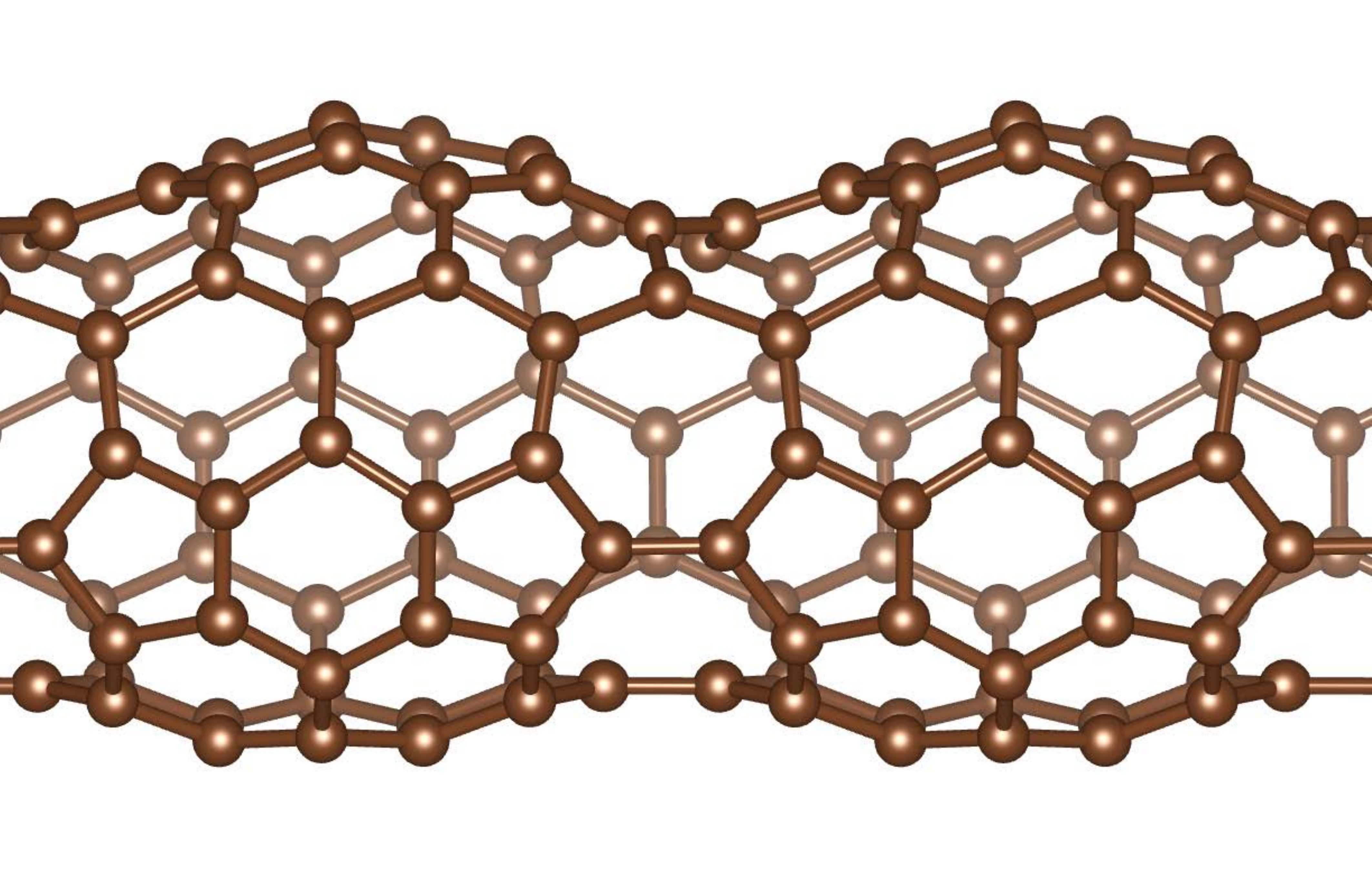}
\includegraphics[width=0.24\hsize,bb= 0 0 1224 790]{T3.pdf}

(m)
\hskip 0.18\hsize (n)
\hskip 0.18\hsize (o)
\hskip 0.18\hsize (p)

\caption{
The shape of the C$_{60}$ polymers given by the first-principles calculations II:
FP5N (a), FP6L (b), FP4K (c), FP4L (d), FP6O (e), FP4J (f), FP5L (g), FP5K (h), FP6F (i), FP6E (j), FP3F (k), FP6J (l), FP6I (m), FP6C (n), FP6K (o), and T3 (p).
}\label{fg:C60polymer}

\end{center}

\end{figure}

\begin{table*}[htb]
\centering
\caption{Total energy and discrete principal curvature of C$_{60}$-polymer:
}\label{tbl:DPC_full2}
  \begin{tabular}{c|cccccccccc}
 & $G$& $L$& $E_{\mathrm{tot,uc}}$ & $\overline{k_1}$ & $\delta k_1$ & $1/\overline{k_1}$ & $\overline{k_2}$ & 
$\delta k_2$ & $1/\overline{k_2}$ & $\overline{|\Delta k_1|}$ \\
C$_{60}$-P& & (\AA)& (eV/cell)& $(\AA^{-1})$& $(\AA^{-1})$& $(\AA)$& $(\AA^{-1})$& $(\AA^{-1})$& $(\AA)$& (\AA)\\
\midrule
FP5N & $D_{5h}$ & 8.128  & -2.177 & 0.306  & 0.022  & 3.267  & 0.034  & 0.121  & 29.047  & 1.164E-02 \\
FP6L & $C_s$ & 7.978  & -1.408 & 0.302  & 0.021  & 3.316  & 0.027  & 0.114  & 37.231  & 1.696E-02 \\
FP4K & $C_s$ & 8.244  & -1.370 & 0.306  & 0.025  & 3.265  & 0.036  & 0.128  & 27.901  & 1.583E-02 \\
FP4L & $C_{2h}$ & 8.250  & -1.233 & 0.308  & 0.034  & 3.252  & 0.035  & 0.126  & 28.725  & 1.956E-02 \\
FP6O & $C_{2h}$ & 7.998  & -1.029& 0.303  & 0.026  & 3.298  & 0.023  & 0.105  & 43.812  & 1.769E-02 \\
FP4J & $C_s$ & 8.210  & -0.860& 0.308  & 0.031  & 3.244  & 0.040  & 0.136  & 25.038  & 1.860E-02 \\
FP5L & $C_1$ & 8.112  & -0.803& 0.305  & 0.030  & 3.280  & 0.029  & 0.120  & 34.045  & 1.886E-02 \\
FP5K & $C_1$ & 8.110  & -0.797 & 0.304  & 0.026  & 3.287  & 0.031  & 0.121  & 32.769  & 1.662E-02 \\
FP6F & $C_2$ & 7.917  & -0.578 & 0.300  & 0.030  & 3.338  & 0.025  & 0.123  & 40.084  & 2.042E-02 \\
FP6E & $C_1$ & 7.921  & -0.452 & 0.300  & 0.029  & 3.328  & 0.025  & 0.124  & 40.567  & 1.871E-02 \\
FP3F  & $C_s$ & 8.333  & -0.442 & 0.308  & 0.028  & 3.251  & 0.042  & 0.140  & 23.908  & 1.909E-02 \\
FP6J & $C_1$ & 7.936  & -0.313& 0.299  & 0.028  & 3.350  & 0.024  & 0.116  & 41.934  & 2.176E-02 \\
FP6I & $C_1$ & 7.941  & -0.184 & 0.300  & 0.029  & 3.335  & 0.023  & 0.115  & 44.278  & 2.223E-02 \\
FP6C & $C_s$ & 7.855  & -0.090& 0.297  & 0.030  & 3.372  & 0.025  & 0.136  & 40.565  & 1.858E-02 \\
FP6K & $C_1$ & 7.939  & -0.088& 0.300  & 0.026  & 3.331  & 0.023  & 0.115  & 44.000  & 1.971E-02 \\
T3&   $D_{5d}$ & 8.694  & 0.000  & 0.322  & 0.049  & 3.109  & 0.064  & 0.167  & 15.510  & 1.142E-02 \\
\bottomrule
  \end{tabular}

\end{table*}

Here we analyzed the geometrical configurations of the C$_{60}$-polymers, or the PSFP models more shown in Table \ref{tbl:DPC_full2}, under the same calculation conditions as in Table \ref{2tb:Ex50} to clarify the relation between the geometrical properties and the total energy.
They have different connections as shown in Figure \ref{fg:C60polymer}, and thus they have different configurations optimized by the first-principles calculations.  The total energy of the nanocarbons lies between those of T3 and FP5N.
Since we found no correlations among $\delta k_1$, $\overline{k_1}$, and the total energy per unit cell $E_{\mathrm{tot, uc}}$.
Hence we are concerned with the homogeneity of $k_1$.
It is known that there is a natural graph Laplacian in trivalent graphs.
For a function $f$ defined on the vertices on a trivalent graph, the graph Laplacian $\Delta$ is defined by
$$
(\Delta f)(x)=f(x)-\frac{1}{3}(f(x_1)+f(x_2)+f(x_3)),
$$
where $x$ is a centered vertex and $x_i$'s denote its adjacent varieties.
$|\Delta(k_1)|$ shows the degree of the homogeneity for $k_1$ locally.
Then we calculated the average $\overline{|\Delta(k_1)|}$ of $|\Delta(k_1)|$ against carbon atoms for each C$_{60}$-polymer, and compared it with the total energy $E_{\mathrm{tot, uc}}$ of the configurations.

\begin{figure}[H]
\begin{center}
\includegraphics[width=0.70\hsize, bb=0 0 650 458]{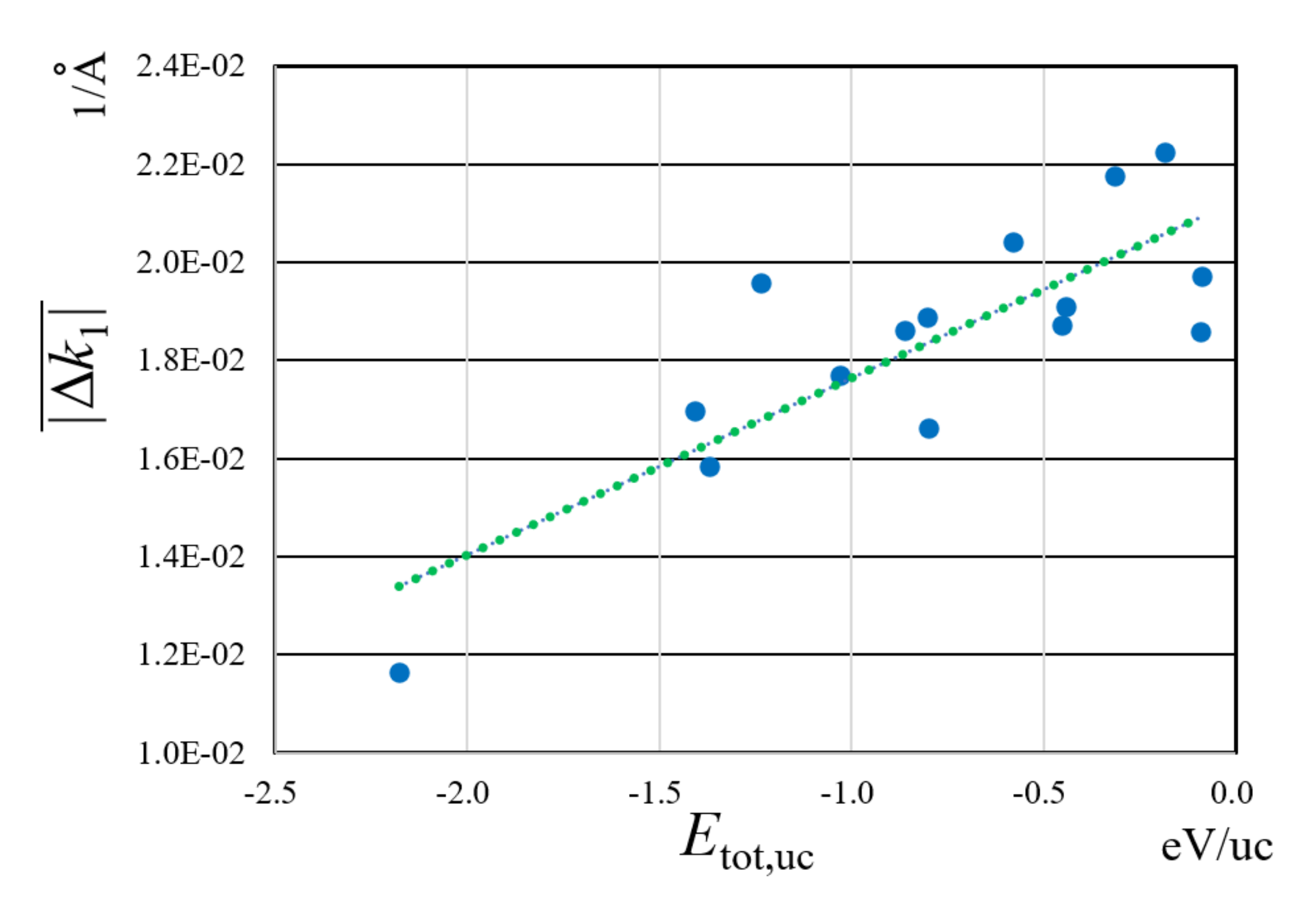}

\caption{
Correlation between $E_{\mathrm{tot, uc}}$ and $\overline{|\Delta k_1|}$:
except $T3$ in Table \ref{tbl:DPC_full2}.
($\overline{|\Delta k_i|} = 0.0036E_{\mathrm{tot, uc}} + 0.0213$, $R^2 = 0.6719$)
}\label{fg:cor_Ek1}

\end{center}

\end{figure}

Figure \ref{fg:cor_Ek1} shows the correlations between $\overline{|\Delta(k_1)|}$  v.s. the total energy $E_{\mathrm{tot, uc}}$ obtained by the first-principles calculations.

Although T3 data were omitted, we found a strong correlations between them as displayed in Figure \ref{fg:cor_Ek1}, which  supports the above depiction.
Accordingly, it can be concluded that the first principal curvature tends to be homogeneous due to the total energy except the relation between $\overline{|\Delta(k_1)|}$ and $E_{\mathrm{tot, uc}}$ of T3 at this stage.
However, it should be emphasized that the pCDPC is revealed to be a novel symmetry of the nanocarbons, which is surely connected with the total energy of the configurations. 

This finding was first obtained by the interdisciplinary study between materials science and pure mathematics.

\bigskip

\section{Conclusion}\label{sec:S7}

Following the first-principles calculations \cite{NOO}, we recomputed the configurations of the carbon atoms in the nanocarbons, C$_{60}$, carbon nanotubes, C$_{60}$ dimer, and the C$_{60}$-polymers.
Since the calculations precisely provide possible configurations of carbons in nanocarbons, we analyzed the geometrical structure of the carbon configurations in the nanocarbons. Then we found that there exists a novel symmetry in the nanocarbons, i.e., the pre-constant discrete principal curvature (pCDPC) structure.

Since the discrete geometry for trivalent oriented graphs was introduced by Kotani, Naito, and Omori \cite{KNO}, we defined the discrete principal curvatures $k_1$ and $k_2$ based on their results.
In terms of the discrete principal curvatures, we numerically investigated the discrete principal curvature distribution of the carbon atom configurations in the nanocarbons by using the data of the first-principles calculations. 
While the C$_{60}$ and nanotubes have the constant discrete principal curvature we expected, as shown in Table \ref{tbl:DPC_full}, it was shown that the C$_{60}$-polymers and C$_{60}$ dimer (C$_{120}$) also have the almost constant discrete principal curvature, i.e., pCDPC, displayed in Figs. \ref{fg:FP5N_dist} and \ref{fg:P08_dist}.
This fact is critical.
The nontrivial pCDPC structure is allowed because of its discrete nature.
Such a revolutionary geometric symmetry has been overlooked in geometry.
In discrete geometry, there appears a center axisoid which is the discrete analogue of the center axis in the continuum differential geometry but has three-dimensional structure rather than a one-dimensional curve due to its discrete nature. 
We showed that such a symmetry exists in nature, namely in the C$_{60}$-polymers, which exist in the real world rather than the fictitious world of mathematics.
Further, we found that there is a strong correlation between the degree of the pCDPC structure and stability of the configurations for certain classes of the C$_{60}$-polymers as in Table \ref{tbl:DPC_full2} and Figure \ref{fg:cor_Ek1}. As argued in the Discussion, it could be explained by the overlap of the C-$2p$ atomic orbitals as illustrated in Figure \ref{fg:Wfunc}. 
It shows the origin of the pCDPC structure in the nanocarbons.

\bigskip

Moreover, the electric system of the C$_{60}$-polymers has significant properties arising from the submanifold quantum system \cite{OISYK}. 
Since the geometrical quantum effects predicted by Jensen and Koppe \cite{JKoppe}, and da Costa \cite{daCosta} was confirmed experimentally \cite{OISYK}, the present study would be extended to the discrete curvature system. 
Namely, the discrete nature of the electric system of the C$_{60}$-polymers becomes important to cultivate new novel nanocarbons. 
For the discrete version of the quantum effects, the path integral method might be proper because the curvature effects appears as the measure of the wave functions and the quantum fields \cite{Ma1,Ma2}.

\subsection*{Acknowledgements:}
This work was supported by Institute of Mathematics for Industry, Joint Usage/Research Center in Kyushu University. 
(``IMI workshop II: Geometry and Algebra in Material Science I'',
September 7--8, 2020", (20200012),
``IMI workshop II: Geometry and Algebra in Material Science II'',
August 30-31, 2021", (20210001),
and
``IMI workshop I: Geometry and Algebra in Material Science III'',
September 8--10, 2022" (2022a003)).
Y. K. has been supported by Grants: Young Scientist of Japan Society for the Promotion of Science Grant, no. JP 20K14312.
S. M. has been supported by the Grant-in-Aid for Scientific Research (C) of Japan Society for the Promotion of Science Grant, No.21K03289.
Y. O. has been supported by Grants: Young Scientist of Japan Society for the Promotion of Science Grant, no. 21K13799.

In the display of the atomic configurations, we used the visualization software, VESTA \cite{MI}.


\newpage

\bibliographystyle{unsrt}
\bibliography{KMNOO2023}

\bigskip

\bigskip

\noindent
Yutaro Kabata\\
School of Information and Data Sciences, \\
Nagasaki University,\\
1-14 Bunkyo-machi, Nagasaki City 852-8131, Japan\\

\bigskip

\noindent
Shigeki Matsutani\\
Electrical Engineering and Computer Science,\\
Graduate School of Natural Science \& Technology, \\
Kanazawa~University,\\
Kakuma Kanazawa, 920-1192, Japan\\
\texttt{s-matsutani@se.kanazawa-u.ac.jp}

\bigskip

\noindent
Yusuke Noda\\
Department of Information and Communication Engineering,\\
Okayama Prefectural University,\\
111 Kuboki, Soja, Okayama 719-1197, Japan\\

\bigskip

\noindent
Yuta Ogata\\
Department of Mathematics, Faculty of Science,\\
Kyoto Sangyo University,\\
Motoyama, Kamigamo, Kita-ku, Kyoto 603-8555, Japan\\

\bigskip

\noindent
Jun Onoe\\
Department of Energy Science and Engineering\\
Nagoya University,\\
Furo-cho, Chikusa-ku, Nagoya, Aichi 464-8603, Japan\\

\end{document}

%% file: define.tex
\usepackage{color}

\def\RR{{\mathbb R}}

\def\bc{{\mathbf c}}

\def\be{{\mathbf e}}

\def\bn{{\mathbf n}}

\def\bv{{\mathbf v}}

\def\bx{{\mathbf x}}











%% file: KMNOO20230629arXiv.bbl
\begin{thebibliography}{10}

\bibitem{ShimaOnoe}
H.~Shima and S.~Onoe.
\newblock {\em Topology-Induced Geometry and Properties of Carbon
  Nanomaterials}, pages 53--84.
\newblock Springer-Verlag, New York, the role of topology in materials edition,
  2018.

\bibitem{MT}
H.~Terrones and A.~L. Mackay.
\newblock Negatively curved graphite and triply periodic minimal surfaces.
\newblock {\em J. Math. Chem.}, 15:183--195, 1994.

\bibitem{T}
H.~Terrones.
\newblock Curved graphite and its mathematical transformations.
\newblock {\em J. Math. Chem.}, 15:143--156, 1994.

\bibitem{IKNSKA}
M.~Itoh, M.~Kotani, H.~Naito, T.~Sunada, Y.~Kawazoe, and T.~Adschiri.
\newblock New metallic carbon crystal.
\newblock {\em Phys. Rev. Lett.}, 102:055703, 2009.

\bibitem{YTSKMI}
Y.~Yao, J.~S. Tse, J.~Sun, D.~D. Klug, R.~Marto\ifmmode~\check{n}\else
  \v{n}\fi{}\'ak, and T.~Iitaka.
\newblock Comment on ``new metallic carbon crystal''.
\newblock {\em Phys. Rev. Lett.}, 102:229601, Jun 2009.

\bibitem{KNO}
M~Kotani, H.~Naito, and T.~Omori.
\newblock A discrete surface theory.
\newblock {\em Comput. Aided Geom. Design}, 58:24--54, 2017.

\bibitem{KNT}
M~Kotani, H.~Naito, and C.~Tao.
\newblock Construction of continuum from a discrete surface by its iterated
  subdivisions.
\newblock {\em Tohoku Math. J.}, 74:215--227, 2022.

\bibitem{S}
T.~Sunada.
\newblock Lecture on topological crystallography.
\newblock {\em Japan. J. Math.}, 7:1--39, 2012.

\bibitem{ONAH}
J.~Onoe, T.~Nakayama, M.~Aono, and T.~Hara.
\newblock Structural and electronic properties of an electron-beam irradiated
  {C}$_{60}$ film.
\newblock {\em Appl. Phys. Lett.}, 82:595--59, 2003.

\bibitem{UONIO}
S.~Ueda, K.~Ohno, Y.~Noguchi, S.~Ishii, and J.~Onoe.
\newblock Dimension dependence of the electronic structure of fullerene
  polymers.
\newblock {\em J. Phys. Chem. B}, 110:22374--22381, 2006.

\bibitem{OISYK}
J.~Onoe, T.~Ito, H.~Shima, H~Yoshioka, and S.~Kimura.
\newblock Observation of riemannian geometric effects on electronic states.
\newblock {\em Eur. Phys. Lett.}, 98:27001 (5pages), 2012.

\bibitem{BO}
T.~A. Beu and J.~Onoe.
\newblock First-principles calculations of the vibrational spectra of one
  dimensional {C}${}_{60}$ polymers.
\newblock {\em Phys. Rev. B}, 74:195426, 2006.

\bibitem{NOO}
Y.~Noda, S.~Ono, and K.~Ohno.
\newblock Geometry dependence of electronic and energetic properties of
  one-dimensional peanut-shaped fullerene polymers.
\newblock {\em J. Phys. Chem. A}, 119:3048--30556, 2015.

\bibitem{Vassilev}
V.~M. Vassilev.
\newblock Unduloid-like equilbrium shapes of single-wall carbon nanotubes under
  pressure.
\newblock In {\em Fourteenth International Conference on Geometry,
  Integrability and Quantization June 8-13, 2012, Varna, Bulgaria}, pages
  244--252, 2013.

\bibitem{Xie_etal}
S.~S. Xie, W.~Z. Li, L.~X. Qian, B.~H. Chang, C.~S. Fu, R.~A. Zhao, W.~Y. Zhou,
  and G.~Wang.
\newblock Equilibrium shape equation and possible shapes of carbon nanotubes.
\newblock {\em Phys. Rev. B}, 54:16436--16439, 1996.

\bibitem{KMO}
Y.~Kabata, S.~Matsutani, and Y.~Ogata.
\newblock On discrete constant principal curvature surfaces.
\newblock {\em arXiv:2306.15846}, 2023.

\bibitem{TON}
A.~Takashima, J.~Onoe, and T.~Nishii.
\newblock In situ infrared spectroscopic and density-functional studies of the
  cross-linked structure of one-dimensional {C}$_{60}$ polymer.
\newblock {\em J. Appl. Phys.}, 108:033514 (5pages), 2010.

\bibitem{MOY}
H.~Masuda, J.~Onoe, and H.~Yasuda.
\newblock High-resolution transmission electron microscopic and electron
  diffraction studies of {C}$_{60}$ single crystal films before and after
  electron-beam irradiation.
\newblock {\em Carbon}, 81:842, 2015.

\bibitem{MYO}
H.~Masuda, H.~Yasuda, and J.~Onoe.
\newblock Two-dimensional periodic arrangement of one-dimensional polymerized
  {C}$_{60}$ evidenced by high-resolution cryo-transmission electron
  microscopy.
\newblock {\em Carbon}, 96:316--19, 2016.

\bibitem{WH}
G.~Wang, Y.~Li, and Y.~Huang.
\newblock Structures and electronic properties of peanut-shaped dimers and
  carbon nanotubes.
\newblock {\em J. Phys. Chem. B}, 109:10957--10961, 2005.

\bibitem{NO}
Y.~Noda and K.~Ohno.
\newblock Metallic and non-metallic properties of one-dimensional peanut-shaped
  fullerene polymers.
\newblock {\em Synth. Met.}, 161:1546--1551, 2011.

\bibitem{KF}
G.~Kresse and J.~Furthm\"uller.
\newblock Efficient iterative schemes for ab initio total-energy calculations
  using a plane-wave basis set.
\newblock {\em Phys. Rev. B}, 54:11169--11186, 1996.

\bibitem{HK}
P.~Hohenberg and W.~Kohn.
\newblock Inhomogeneous electron gas.
\newblock {\em Phys. Rev.}, 136:B864--B871, 1964.

\bibitem{PBE}
J.~P. Perdew, K.~Burke, and M.~Ernzerhof.
\newblock Generalized gradient approximation made simple.
\newblock {\em Phys. Rev. Lett.}, 77:3865--3868, 1986.

\bibitem{KJ}
G.~Kresse and D.~Joubert.
\newblock From ultrasoft pseudopotentials to the projector augmented-wave
  method.
\newblock {\em Phys. Rev. B}, 59:1758--1775, 1999.

\bibitem{JKoppe}
H.~Jensen and H.~Koppe.
\newblock Quantum mechanics with constraints.
\newblock {\em Ann. Phys.}, 63:586--591, 1971.

\bibitem{daCosta}
R.~C.~T. da~Costa.
\newblock Quantum mechanics of a constrained particle.
\newblock {\em Phys. Rev. A}, 23:1982--7, 1981.

\bibitem{Ma1}
S.~Matsutani.
\newblock Path integral formulation of curved low dimensional space.
\newblock {\em J. Phys.Soc.Jpn.}, 61:53--63, 1992.

\bibitem{Ma2}
S.~Matsutani.
\newblock Quantum field theory on curved low dimensional space embedded in
  three dimensional space.
\newblock {\em Phys. Rev. A}, 47:686--689, 1993.

\bibitem{MI}
K.~Momma and F.~Izumi.
\newblock Vesta 3 for three-dimensional visualization of crystal, volumetric
  and morphology data.
\newblock {\em J. Appl. Crystallogr.}, 44:1272--1276, 2011.

\end{thebibliography}
